\documentclass[journal=jctcce,layout=onecolumn]{achemso}

\usepackage[version=3]{mhchem} 
\usepackage{gensymb}
\usepackage[font=small]{caption}

\newcommand{\fig}[1]{Fig.~\ref{fig:#1}~} 

\newcommand*{\tab}[1]{Table~\ref{tab:#1}}

\newcommand{\cn}[1]{$^\textnormal{\tiny\color{red}cite}$}

\RequirePackage{color}\definecolor{chris}{rgb}{0,.5,.5}
\newcommand{\SM}{{Sm$^{4+}$}}

\newcommand{\MG}{{Mg$^{2+}$}}

\newcommand{\NA}{{Na$^+$}}
\newcommand{\LI}{Li$^+$}
\newcommand{\K}{K$^+$}

\newcommand*{\rmin}{r_\textnormal{min}}
\newcommand*{\rminNOC}{r_\textnormal{min}^{\textnormal{N}-\textnormal{O}=\textnormal{C}}}
\newcommand*{\sigNOC}{\sigma^{\textnormal{N}-\textnormal{O}=\textnormal{C}}}
\newcommand*{\rminNOS}{r_\textnormal{min}^{\textnormal{N}-\textnormal{O}=\textnormal{S}}}
\newcommand*{\rminNOP}{r_\textnormal{min}^{\textnormal{N}-\textnormal{O}=\textnormal{P}}}
\newcommand*{\sigNOP}{\sigma^{\textnormal{N}-\textnormal{O}=\textnormal{P}}}

\newcommand*{\mytitle}{Improved Parameterization of Amine--Carboxyate and Amine--Phosphate Interactions for Molecular Dynamics Simulations Using the CHARMM and AMBER Force Fields}

\renewcommand*{\fig}[1]{Figure~\ref{fig:#1}}
\author{Jejoong Yoo}
\affiliation[Department of Physics]{Center for the Physics of Living Cells, Department of Physics, University of Illinois at Urbana--Champaign, 1110 West Green Street, Urbana, Illinois 61801}

\author{Aleksei Aksimentiev}
\email{aksiment@illinois.edu}
\affiliation[Department of Physics]{Center for the Physics of Living Cells, Department of Physics, University of Illinois at Urbana--Champaign, 1110 West Green Street, Urbana, Illinois 61801}

\title{\mytitle}


\begin{document}


\clearpage
\begin{abstract}  
    Over the past decades, molecular dynamics (MD) simulations of biomolecules have become a mainstream biophysics technique. As the length and time scales amenable to the MD method increase, shortcomings of the empirical force fields---which have been developed and validated using relatively short simulations of small molecules---become apparent. One common artifact is aggregation of water-soluble biomolecules driven by artificially strong charge--charge interactions.  Here, we report a systematic refinement of Lennard-Jones parameters (NBFIX) describing amine--carboxylate and amine--phosphate interactions, which brings MD simulations of basic peptide-mediated nucleic acids assemblies and lipid bilayer membranes in better agreement with experiment.
%
    As our refinement neither affects the existing parameterization of bonded interaction nor does it alter the solvation free energies, it improves realism of an MD simulation without introducing additional artifacts.

\end{abstract}

%
%
%
%
%
%

\clearpage
\section{Introduction}


Among common chemical groups that define the structure and physical properties of biomacromolecules, charged chemical moieties bear particular importance as they guide assembly of chemically distinct molecules into functional  biological units, whether it is binding of a neurotransmitter to an ion channel, assembly of a ribosome or association of peripheral proteins with a lipid bilayer membrane.  
In proteins, the charged moieties are amine-based (lysine and arginine) and carboxylate-based (aspartate and glutamate) side chains, that
 carry positive and negative charges, respectively, in aqueous environment and at physiological pH.
In a nucleic acid polymer, every nucleic acid monomer contains a negatively charged phosphate group, which gives the polymer its high negative charge.
%
%
In lipids, anionic head groups are formed by  negatively charged carboxylate- or phosphate-based moieties, 
whereas polar  head groups contain both  negatively charged phosphates and positively charged amines.
%
%
%
Thus, non-bonded interactions between positively charged amines and negatively charged carboxylates or phosphates  are central to association of  proteins, nucleic acids, and lipid bilayer membranes into functional biological assemblies. 

In the case of protein--DNA assemblies, interactions between basic residues of the protein and the phosphate groups of DNA backbone or the electronegative sites of DNA bases are essential for sequence-specific binding of transcription factors via leucine zipper, zinc finger, helix-turn-helix mechanisms~\cite{PABO1984,PABO1992}.
In helicases, basic residues of the DNA-binding domains bind the phosphate groups of the DNA backbone to  transmit
the mechanical force generated by the helicase' motor domain to the DNA~\cite{THOM2009}.
In nucleosomes, binding of arginine and lysine residues  to DNA  stabilizes  the protein--DNA complex~\cite{LUGE1997,BINT2012}.  Amine--phosphate interactions were found to modulate the rate  of electrophoretic transport of single-stranded DNA through biological nanopores~\cite{RINC2011,BHAT2012}. 

Formation of salt bridges between amine and carboxylate groups is also important for protein stability~\cite{PERU1978,SHEI2000}.
For example, a salt bridge between lysine and aspartate residues is essential for the stability of  $\beta$-amyloid fibrils implicated in the Alzheimer disease~\cite{PETK2002}. Salt bridges between lysine and aspartate side chains are critical to structural stability and ligand binding behavior of the extracellular domain of G-protein coupled receptors (GPCR)~\cite{JI1998,BOKO2010}.


In protein--lipid systems, charged amino acids maintain the 
structure~\cite{kYTE1982,HESS2005,VONH2006}
and enable biological function~\cite{SCHM2006,MCLA2005} of membrane proteins. 
 Charged residues  at the outer surface of an integral membrane protein stabilize its placement in a lipid bilayer membrane through interactions with the lipid head groups~\cite{VONH2006,MCLA2005}.
One exemplary system is a voltage-gated potassium channel, where arginine--lipid phosphate interactions are essential for the biological function of the channel~\cite{SCHM2006}.
According to molecular dynamics (MD) simulations, such arginine--phosphate interactions can also affect organization of lipid molecules surrounding the protein~\cite{FREI2005}.
%
%
%
Amine--carboxylate interactions are also essential for the function of ligand-gated ion channels~\cite{WANG2015}. 
In fact, the majority of neurotransmitters contain amine and carboxylate moieties, including AMPA ($\alpha$-amino-3-hydroxy-5-methyl-4-isoxazole propionic acid), NMDA (N-methyl-D-aspartate), GABA ($\gamma$-aminobutyric acid), and nAChR (nicotinic acetylcholine)~\cite{BETZ1990,ARMS1998}.
The the binding sites of  ligand-gated ion channels feature complementary patterns of charged amino-acids 
that coordinate ligand binding via   amine--carboxylate interactions~\cite{BETZ1990,ARMS1998}.

Over the past decades, MD simulations have provided a wealth of information about all of the  above  systems. The accuracy of the MD method has been gradually increasing thanks to many parametrization efforts, in particular refinement of the bonded interactions describing  proteins~\cite{CORN1995,HORN2006,BUCK2006,BEST2012,LIND2010B}, lipids~\cite{KLAU2005,KLAU2010}, and nucleic acids~\cite{CORN1995,CHARMMDNA2,FOLO2000,PERE2007,HART2012,DENN2011,ZGAR2011B,FADR2009}.
Much less efforts, however, were devoted to refinement of non-bonded interactions.
In the CHARMM force field, parameters describing the  non-bonded  interactions of amine, carboxylate, and phosphate groups were developed in 1998 by matching  the solute--water interaction energies against  the results of quantum mechanics calculations and experiment~\cite{MACK1998}. 
These parameters survived all subsequent revisions of the CHARMM force field without any modifications.
Similarly, all variants of the AMBER force fields rely on non-bonded parameterization of amine--carboxylate and amine--phosphate interactions  developed in 1995~\cite{CORN1995}.

A potential concern with relying on conventional parameterization of amine--carboxylate and amine--phosphate interactions is that these interactions have never been explicitly calibrated or validated during the force field development process~\cite{MACK1998,CORN1995}.
Both CHARMM and AMBER force fields were parameterized to reproduce individual water--solute interactions, which may not be sufficient 
to describe interactions between two solvated and oppositely charged chemical groups.
It is not, perhaps, surprising that, as the time and length scales of the MD method increase,  artifacts related to inaccuracies of non-bonded parameterization become to emerge.
For example, simulations of villin head piece proteins in aqueous environment suffer from artificial aggregations in both CHARMM and AMBER models~\cite{PETR2014}. Artificial aggregation has also been reported for CHARMM and AMBER simulations of 
 short peptides~\cite{JOHN2009,NERE2012,BEST2014,PIAN2015}.


 

Recently, we have shown that the two most popular empirical force fields (AMBER and CHARMM) considerably overestimate 
attractive forces between cations, e.g., \LI, \NA, \K, and \MG,  and anionic moieties , e.g., acetate and phosphate~\cite{YOO2012}.
 In MD simulations of DNA assemblies, these force field artifacts lead to artificial  aggregations of DNA helices~\cite{YOO2012}, which are experimentally known to repel one another at identical conditions~\cite{RAU1984}.
%
Thus, a seemingly minor imprecision in the description of non-bonded interactions between select atom pairs can have disastrous consequences on the large scale behavior of a molecular system, leading to qualitatively wrong simulation outcomes.
We have also shown~\cite{YOO2012} that surgical refinement~\cite{LUO2009b}  of a small set of non-bonded parameters against experimental osmotic pressure data dramatically increases the realism of DNA array simulations, bringing them in quantitative agreement with experiment~\cite{YOO2012B,MAFF2014,YOO2015A}.

In this work, 
we use solutions of glycine, ammonium sulfate, and DNA to evaluate and improve parameterization of  amine--carboxylate and amine--phosphate interactions within the AMBER and CHARMM force fields.
Through comparison of the simulated and experimental osmotic pressure data, we demonstrate that both AMBER and CHARMM force fields significantly overestimate attraction between solvated amine and carboxylate/phosphate groups.
To remedy the problem, we develop a set of corrections to the van der Waals (vdW) parameters describing non-bonded interactions between specific nitrogen and oxygen atom pairs (NBFIX). We show that using our corrections brings the results of MD simulations of model compounds in quantitative agreement with experiment. 
Further, we detail the effect of our NBFIX corrections on MD simulations of  lysine-mediated DNA--DNA forces and  lipid bilayer membranes, concluding that our corrections improve overall realism of MD simulations without introducing additional artifacts.

\section{Methods}

\subsection{Empirical force fields}


All our MD simulations based on the CHARMM force field employed its most recent variant (CHARMM36), including the updated parameters for DNA~\cite{HART2012}, lipids~\cite{KLAU2010} and CMAP corrections~\cite{BEST2012}.
For our lipid bilayer membrane simulations only, we also used the CHARMM27r parameter set to compare with CHARMM36~\cite{KLAU2005}.
For water, we employed the TIP3P model~\cite{JORG1983} modified for the CHARMM force field~\cite{PRIC2004}.
For ions, we employed the standard CHARMM parameters~\cite{BEGL1994} and our custom NBFIX corrections to describe  ion-ion, ion-carboxylate, and ion-phosphate interactions~\cite{YOO2012}.
For sulfate and ammonium ions, CHARMM-compatible models were taken from Refs.~\cite{CANN1994} and \cite{ULLM2012}, respectively.

All our MD simulations based on the AMBER force field used its AMBER99 variant. 
Amino acids and proteins were described using the AMBER99sb-ildn-phi parameter set~\cite{LIND2010,NERE2011}, which was shown
to be optimal  for  MD simulations of short peptides and small proteins~\cite{BEAU2012}.
DNA was described using the AMBER99bsc0 parameter set~\cite{PERE2007}.
Lipid bilayer membranes were not simulated using the AMBER force field.
For water, we employed the original TIP3P model~\cite{JORG1983}.
For ions, we employed the ion parameters developed by the Cheatham group~\cite{JOUN2008} and our custom NBFIX corrections for the description of  ion-ion, ion-carboxylate, and ion-phosphate interactions~\cite{YOO2012}.

\subsection{MD simulation of osmotic pressure}

All MD simulations of glycine and ammoniumsulfate solutions were performed using the NAMD2 package~\cite{PHIL2005}
following a previously described protocol~\cite{LUO2009,YOO2012}.
 Each simulation system  contained two compartments separated
  by two virtual semipermeable membranes aligned with the $xy$ plane.  
 One compartment   contained an electrolyte solution whereas the other contained pure water.
The semipermeable membranes were modeled as half-harmonic planar potentials that acted only on the solute molecules and not on water:
	\begin{equation} \label{eqn:memb}
		F^{\textnormal{ memb}}_i = \left\{ \begin{array}{lcl} 
			-k ( z_i - D/2) & \mbox{for} &  z_i > D/2 \\
							0	 & \mbox{for} & |z_i| \le D/2 \\
			-k ( z_i + D/2) & \mbox{for} &  z_i < -D/2 \\
		\end{array}\right.	
	\end{equation}
	where $z_i$ is the $z$ coordinate of solute $i$, $D$ is the width of the electrolyte compartment, and the force constant $k =$ 4,000 kJ/mol$\cdot$nm$^2$~\cite{YOO2012}.
The tclBC function~\cite{HENG2006} of the NAMD package was used to enforce such half-harmonic potentials on the solute molecules.
 
For a given simulation condition, such as the concentration of solutes and/or a particular value of the  interaction parameter, we first
equilibrated the system for at least 1 ns, which was followed by a production simulations of $\sim$25~ns or more.
During each production simulation, we recorded  the instantaneous forces applied by the confining potentials to the solutes,  which were equal by magnitude and opposite in direction to the forces applied by the solutes to the membrane.
 The instantaneous  pressure on the membranes was obtained by dividing the instantaneous total force on the membranes by the total area of the membranes.
The osmotic pressure of a specific system was computed by averaging the instantaneous pressure values.
The  statistical error in the determination of an osmotic pressure value was characterized as the standard deviation of the 1-ns block averages of the instantaneous pressure data. 
 Further details of our simulation protocol can be found in Ref.~\cite{YOO2012}.

All vdW forces were evaluated using a 10-12~\AA~switching scheme. 
 Long-range electrostatic interactions were computed using the Particle Mesh Ewald (PME) method with a 1~\AA~grid spacing. 
The PME calculations were performed every step; a multiple time stepping scheme was not used.
Temperature was kept constant at 298~K using the Lowe-Andersen thermostat~\cite{KOOP2006}.
The area of the systems within the plane of the semi-permeable membranes was kept constant;  the barostat target for the system size fluctuations along the $z$-axis  (perpendicular to the membranes) was set to 1~bar.
The integration time step was 2~fs.

\subsection{MD simulations of DNA array}

All MD simulations of DNA arrays were performed in a constant temperature/constant pressure ensemble (300 K and 1 bar) using  the Gromacs 4.5.5 package~\cite{HESS2008} and the  integration time step of 2~fs. The temperature and pressure were controlled using the Nos\'e-Hoover~\cite{NOSE1983,HOOV1985} and the Parrinello-Rahman~\cite{PARR1981}  schemes, respectively.
A 7-to-10~\AA\  switching scheme was used to evaluate the vdW forces. The long-range electrostatic forces were computed using the PME summation scheme~\cite{DARD1993} with a 1.5~\AA~grid spacing and a 12~\AA~real-space cutoff. Covalent bonds to hydrogen atoms in water and in DNA or spermine were constrained using the SETTLE~\cite{MIYA1992} and LINCS~\cite{HESS1997} algorithms, respectively.

A system containing an array of 64 double stranded (ds) DNA molecules at a neutralizing concentration of {\SM}, \ce{NH3(CH2)3NH2(CH2)4NH2(CH2)3NH3}, was build as follows.
 A  20-bp dsDNA molecule (dG$_{20}$$\cdot$dC$_{20}$), effectively infinite under the periodic boundary conditions, was equilibrated for
   $\sim$30~ns in the presence of 10 {\SM} molecules and explicit water. The simulation unit cell measured $\sim2.5\times2.5\times6.8$~nm$^3$.
The microscopic conformation observed at the end of the equilibration simulation was replicated 64 times to obtain an array of 64 dsDNA molecules.  The replica systems were arranged  within an 11-nm radius disk, avoiding overlaps between the systems as much as possible.  
%
A pre-equilibrated  volume of water was then added to make a rectangular system  that measured $\sim25\times25\times6.8$~nm$^3$.
%
The resulting system  was equilibrated for $\sim$50~ns in a constant temperature/constant pressure ensemble (300 K and 1 bar). During the equilibration, the half-harmonic restraining potential (see the next paragraph) kept DNA inside the cylindrical volume of 11-nm radius.


%
%

To measure the internal pressure of the DNA array, we confined the DNA array by a cylindrical half-harmonic potential~\cite{YOO2012}: 
	\begin{equation} \label{eqn:array}
		V_{\textnormal{array}}(r)  = \left\{ \begin{array}{lcl} 
			 \frac{1}{2} k (r - R)^2 & \mbox{for} &  r > R \\
			  0	& \mbox{for} &   r \le R   
		\end{array}
		\right.	
	\end{equation}
where $r$ was the axial distance from the center of the DNA array, $R = 12.5$~nm was the radius of the confining cylinder, and $k$ was the force constant (100 kJ/mol$\cdot$nm$^2$). 
The restraining potential  applied only to DNA phosphorus atoms.
The radius of the restraining potential, $R=12.5$~nm, was chosen to give the average inter-DNA distance of $\sim$28~\AA~to an 
array of 64 dsDNA molecules arranged on an ideal hexagonal lattice.
Experimentally,  the equilibrium inter-DNA distance in a DNA condensate at 2-mM {\SM} and zero external pressure is $\sim$28~\AA~\cite{TODD2008}. Thus, our simulation setup reproduces experimental conditions where stress-free DNA condensate is observed.

\subsection{MD simulations of DNA--DNA interactions}

All MD simulations of the DNA--DNA PMF were performed in a constant temperature/constant pressure ensemble (300 K and 1 bar) using  the Gromacs 4.5.5 package~\cite{HESS2008}.  With the exception of the restraining potential, all other simulation parameters were same as in our simulations of the DNA array system (see above).

To characterize the free energy of inter-DNA interaction, we computed the DNA--DNA potential of mean forces (PMF) as a function of the inter-axial distance between two dsDNA molecules using the umbrella sampling and weighted histogram analysis (WHAM) methods~\cite{KUMA1992}.
The systems used for the calculations of the DNA--DNA PMF were prepared by 
placing a pair of identical  21 base pair (bp) dsDNA molecules parallel to each other and the $z$-axis of the simulation system.
The sequence of each strand of each dsDNA molecule was 5$'$-(dGdC)$_{10}$dG-3$'$.  Following that, four different systems were made by randomly adding 230 lysine monomers (Lys$^{+}$), 49  lysine dimers (Lys$^{2+}$), 30 lysine trimers (Lys$^{3+}$), or 21 lysine tetramers (Lys$^{4+}$) to the system, corresponding to bulk concentrations of [Lys$^{+}$] $\sim$ 200~mM, [Lys$^{2+}$] $\sim$ 10~mM, [Lys$^{3+}$] $\sim$ 2~mM, and [Lys$^{4+}$] $\sim$ 0.5~mM, respectively. 
Each system was then solvated with $\sim$35,000 TIP3P water molecules and neutralized by replacing 146, 14, 6, or 0 water molecules, respectively,  with chloride ions.
Each system was subject to the hexagonal prism periodic boundary conditions defined by the following unit cell parameters: $a = b = \sim13$~nm, $c=7.14$~nm, $\alpha=\beta=90$\degree, and $\gamma=60$\degree.  The 5$'$ and 3$'$ ends of each strand were covalently linked to each other across the periodic boundary of the unit cell, which made the DNA molecules effectively infinite. 

Each  of the three systems was equilibrated for at least 10~ns in the absence of any restraints.
For each system, the umbrella sampling simulations began from the molecular configuration obtained at the end of the respective equilibration simulation. 
The inter-axial distance---defined as the distance between the centers of mass of the DNA molecules  projected onto the $xy$-plane---was used as a reaction coordinate. Among the umbrella sampling systems,  the  inter-axial distance varied from  25 to 35~\AA\ in  1~{\AA} increments. 
The force constant of the harmonic umbrella potentials was $2,000$ kJ/mol$\cdot$nm$^2$.
For each window, we performed a 10~ns equilibration followed by a $\sim$40~ns production run.
 The PMF was reconstructed using WHAM~\cite{KUMA1992}.

\subsection{MD simulations of lipid bilayer membranes}

 All MD simulations of  lipid bilayer systems were performed using the C37a2 version of the CHARMM package featuring the domain decomposition (DOMDEC) algorithm~\cite{BROO2009b,HYNN2014}.
To be consistent with the standard procedures of the CHARMM force field development, we used the CHARMM input file---generously provided by Dr. Klauda---to specify the parameters of the non-bonded forces calculations as well as the temperature- and pressure-coupling schemes~\cite{KLAU2010}.
For the initial coordinates, we used the pre-equilibrated  lipid bilayer systems  containing 80 POPE, 72 DPPC, 72 DOPC, or 72 POPC lipid molecules~\cite{KLAUDA}.
The POPE system was used without modifications. 
The DPPE, DOPS, and POPS systems were prepared by modifying the lipid head groups of the pre-equilibrated DPPC, DOPC, and POPC systems, respectively, using the internal coordinate manipulation commands (IC) of the CHARMM package.
All simulations were performed in a constant temperature/constant tension ensemble (NP$\gamma$T). 
The tension was always set to zero;  the temperature was adjusted to match the experimental target data.
A 8-12~\AA~force-switching scheme was used to evaluate the vdW forces.
The long-range electrostatic forces were computed using the PME method~\cite{DARD1993} with a 1~\AA~grid spacing.
 The integration time step was 2~fs.
For the PE lipid systems, no ions were added to the solution.
For the PS lipid systems, 72 \NA~ions were randomly placed in the solution to neutralize the system.
The interactions of \NA with the lipid membrane were described using our optimized LJ parameters for \NA--acetate and \NA--phosphate oxygen pairs~\cite{YOO2012}.

\section{Results and discussion}

In an empirical MD force field,  the strength of non-bonded interaction between two solvated chemical groups is affected by the partial electrical charges of the atoms comprising the groups, atom-type specific vdW force constants and the properties of the solvent model.
For our refinement of amine--carboxylate and amine--phosphate interactions, we will alter the atom pair-specific Lennard-Jones parameters leaving the partial charges and water model unchanged. Doing so reduces the possibility of introducing  uncontrolled  artifacts by the 
force field refinement~\cite{LUO2009b}.

\begin{table}[t!]
	\caption{
	Optimized values of the Lennard-Jones (LJ) $\rmin$ (CHARMM) and  $\sigma$ (AMBER) parameters for the description of the vdW interaction between the nitrogen atom of an amine  group and the oxygen (O) atom of a carboxylate or phosphate group.  These parameters were optimized for use with  CHARMM22, CHARMM27, CHARMM36, and all AMBER99-derived force fields. 
	For CHARMM, the table lists 
	the standard CHARMM LJ $\rmin$ parameters ($\rmin^\mathrm{CHM}$), 
	pair-specific adjustments to the LJ $\rmin$ parameters ($\Delta \rmin$),
	and the optimized values of the LJ  $\rmin$ parameter ($\rmin^\mathrm{NBFIX}$).
	For AMBER, the table lists 
	the standard AMBER LJ $\sigma$ parameters ($\sigma^\mathrm{AMB}$), 
	pair-specific adjustments to the LJ $\sigma$ parameters ($\Delta \sigma$),
	and the optimized values of the LJ  $\sigma$ parameter ($\sigma^\mathrm{NBFIX}$). 
	Note that $\rmin = 2^{1/6} \sigma$ for the $6-12$ LJ potential.
	All $\rmin$ and $\sigma$ values are given in  \AA.
	}
	\label{tab:nbfix}
	\begin{tabular}{lccccccc}
		\hline
		& \multicolumn{3}{c}{CHARMM} & & \multicolumn{3}{c}{AMBER} \\
		\cline{2-4}
		\cline{6-8}
		Atom pair & $\rmin^\mathrm{CHM}$  & $\Delta \rmin$   & $\rmin^\mathrm{NBFIX}$  &
		&
		$\sigma^\mathrm{AMB}$  & $\Delta \sigma$   & $\sigma^\mathrm{NBFIX}$  \\
		\hline
		N--O$=$C  & 3.55 & 0.08 & 3.63 & & 3.10 & 0.08 & 3.18 \\ 
		N--O$=$P  & 3.55 & 0.16 & 3.71 & & 3.10 & 0.14 & 3.24 \\ 
		\hline
	\end{tabular}
\end{table}

\subsection{Calibration of amine--carboxylate interactions through simulations of glycine solutions}

\begin{figure}[t!]
    \center
    \includegraphics[width=6.25in]{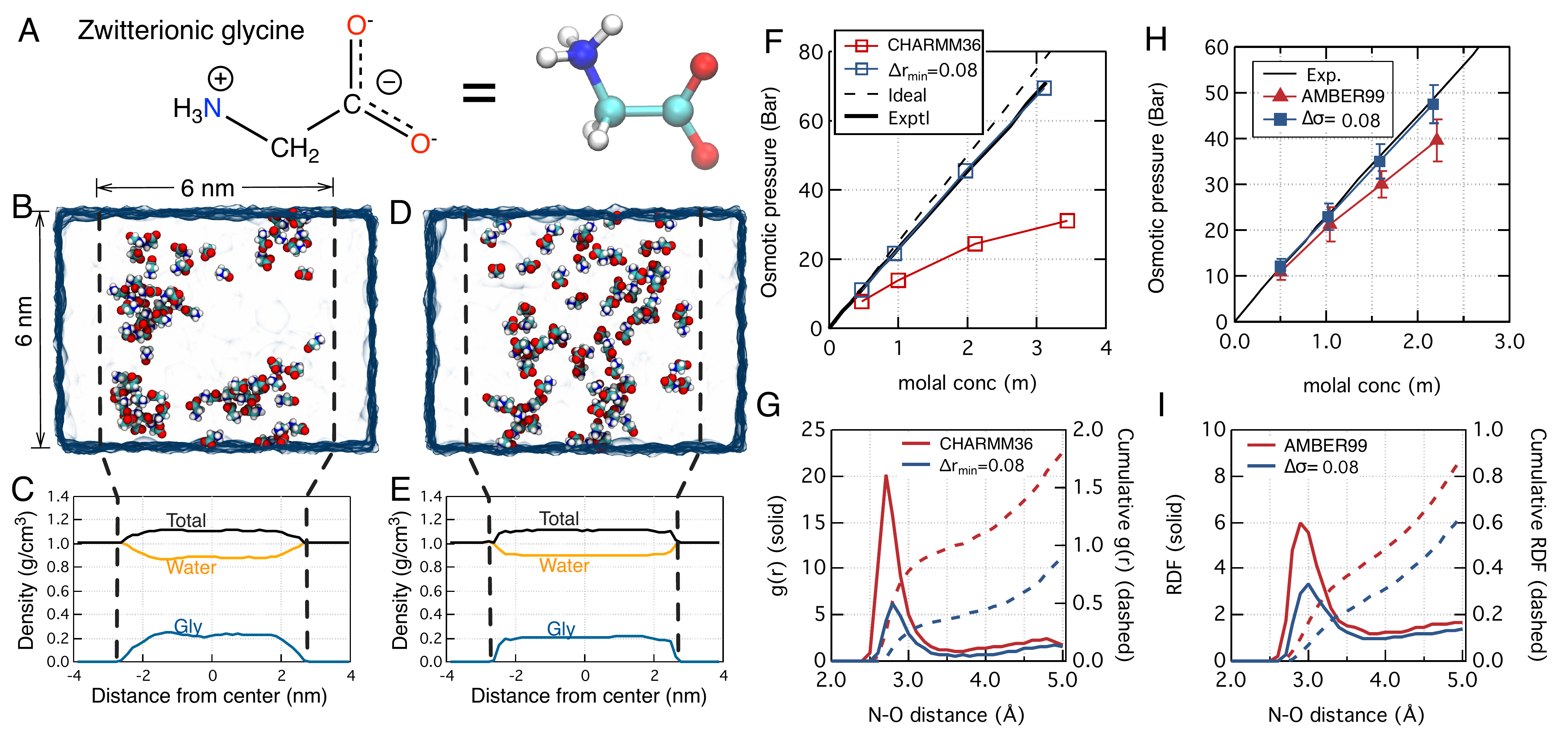}
    \caption{\label{fig:gly} 
   Calibration simulations of glycine monomer solutions.
    (A) Chemical structure of glycine. 
	The oxygen and nitrogen atoms subject to the NBFIX correction are shown in red and blue, respectively.
	  The NBFIX values are given in \tab{nbfix}.
    (B) A simulation system used for calibration simulations. The simulation unit cell contains water (blue semi-transparent surface) 
    divided into two compartments by two planar half-harmonic potentials (depicted by dashed lines).
 One of the compartments contains 
  a  $\sim3$~m aqueous solution of glycine (shown as vdW spheres), the other only pure water.  The half-harmonic potentials apply restraining forces to the atoms of glycine molecules confining them to the same compartment;  the potentials do not act on water molecules. The force exerted by the potentials on the glycine molecules reports on the osmotic pressure of glycine in the glycine-filled compartment. The snapshot illustrates the system's configuration observed
    at the end of a 25-ns simulation performed using  the standard CHARMM36 force field.
    For clarity, only a 10~\AA-wide slice of the system is shown. 
    (C) Density profiles of water and glycine in the direction normal to the planes of the confining potentials for the system simulated using the standard CHARMM36 force field. Each data point represents an average over a $\sim$1.5-\AA-slice of the system and a $\sim$25-ns MD trajectory.
    (D,E) Same as in panels B and C but for the simulations performed applying our NBFIX correction.
    (F) Osmotic pressure of a glycine solution as a function of the glycine concentration  obtained from MD simulations performed  using CHARMM36  with (blue) and without (red) our NBFIX corrections.  For comparison, experimentally determined~\cite{TSUR2007} and ideal solution (osmotic coefficient = 1) dependences of the osmotic pressure
    on solute concentration are shown as solid and dashed black lines, respectively.
    (G) Radial distribution functions, $g(r)$ (solid lines), and cumulative $g(r)$ (dashed lines) for a $\sim$3~m glycine solution simulated using  CHARMM36  with (blue) and without (red) our NBFIX correction. 
	All N--O pairs within the same glycine molecules were excluded from the calculations of $g(r)$.
    (H,I) same as in panels F and G but for the simulations performed using the AMBER99 force field with (blue) and without (red) our NBFIX corrections.
    }
\end{figure}

An aqueous solution of glycine monomers is a convenient system for calibration of  the amine--carboxylate interaction. At physiological pH, glycine monomers prevalently adopt a zwitterionic form containing both charged amine and  charged carboxylate groups in the same 
molecule,~\fig{gly}A.  Because direct charge-charge interactions are considerably stronger than any other non-bonded interactions in solution~\cite{MASU2003},  the interaction between two zwitterionic glycine monomers is  dominated by the amine--carboxylate interaction.  
To test and improve the parameterization of  the amine--carboxylate interaction, we compare the experimentally measured osmotic pressure of a glycine solution to the value obtained by the MD method.  In the case of discrepancy, we modify the parameterization of the   amine--carboxylate interaction until agreement between the simulated and experimentally measured osmotic pressure is reached. 

\fig{gly}B  illustrates a simulation system used to measure the osmotic pressure  of  a glycine solution by the MD method~\cite{LUO2009,YOO2012}. 
	Two virtual semipermeable membranes split the volume of the simulation cell into two compartments, 
one compartment  containing  a glycine solution (the solute compartment) and the other containing pure water (the water compartment).
The membranes, modeled as half-harmonic potentials acting on the solute molecules only, allow water to pass between the compartments while forcing solutes to remain within the solute compartment. 
 The details of the simulation protocol are provided in Methods.
%

\fig{gly}C illustrates a steady-state density distribution in the simulation system. 
The differential partitioning of the molecules among the compartment 
creates osmotic pressure, which, in our simulations, is balanced by the forces applied by the membranes.
At a 3~m concentration, glycine monomers aggregate  in the simulation  performed using the standard CHARMM force field,~\fig{gly}B, whereas in experiment they remain fully soluble~\cite{crc_handbook}. This observation suggests that the standard parameterization of the CHARMM force field overestimates the strength of amine--carboxylate interaction. A quantitative confirmation of the above conclusion comes from the
comparison of the simulated and experimentally measured~\cite{TSUR2007} osmotic pressure at several glycine concentrations,~\fig{gly}F. 
The standard parameterization considerably underestimates the osmotic pressure at all concentrations; the discrepancy between simulation and experiment becomes larger as the glycine concentration  increases.  

To improve the force field model, we gradually increased the Lennard-Jones (LJ) $\rmin$ parameter describing the vdW interaction  of amine nitrogen (N) and carboxylate oxygen (O=C), $\rminNOC$.  Our custom corrections to the $\rmin$ parameters of atomic pairs are introduced in MD simulations using the NBFIX option of the CHARMM force field and hereafter are referred to as NBFIX corrections.
In a series of osmotic pressure simulations of the 3~m glycine solution,
the osmotic pressure increased monotonically with the $\rminNOC$ parameter, reaching the experimental value when increased by 0.08~\AA\ from the standard CHARMM value~($\Delta \rminNOC = 0.08$~\AA),~Fig.~S1A. 
%
Similar improvement was observed over the entire range of glycine concentration,~\fig{gly}F.
In the simulation employing our NBFIX correction, glycine monomers are homogeneously distributed over the solute compartment at 3~m concentration,~\fig{gly}D,~E. The  radial distribution function, $g(r)$, and the cumulative $g(r)$  of the inter-molecular pair of amine nitrogen and carboxylate oxygen atoms  indicate a considerable suppression of the direct contacts when our NBFIX correction is enabled,~\fig{gly}G.

Next, we repeated our procedures to calibrate the amine--carboxylate interactions in the AMBER99 force field. As in the case of CHARMM, the osmotic pressure of a 2~m glycine solution simulated using the standard AMBER force field was significantly smaller than the experimental value,~\fig{gly}H.
We could reproduce the experimental osmotic pressure over the entire range of glycine concentration by increasing the LJ $\sigma$ parameter describing the vdW interaction  of amine nitrogen and carboxylate oxygen ($\sigNOC$) by 0.08~\AA. Using this NBFIX correction
in an MD simulation considerably reduced the number of direct contacts between glycine monomers,~\fig{gly}I.

\subsection{Calibration of amine-sulphate/phosphate interactions for the CHARMM force field through simulations of ammonium sulphate solutions}

\begin{figure}[t!]
    \center
    \includegraphics[width=3.25in]{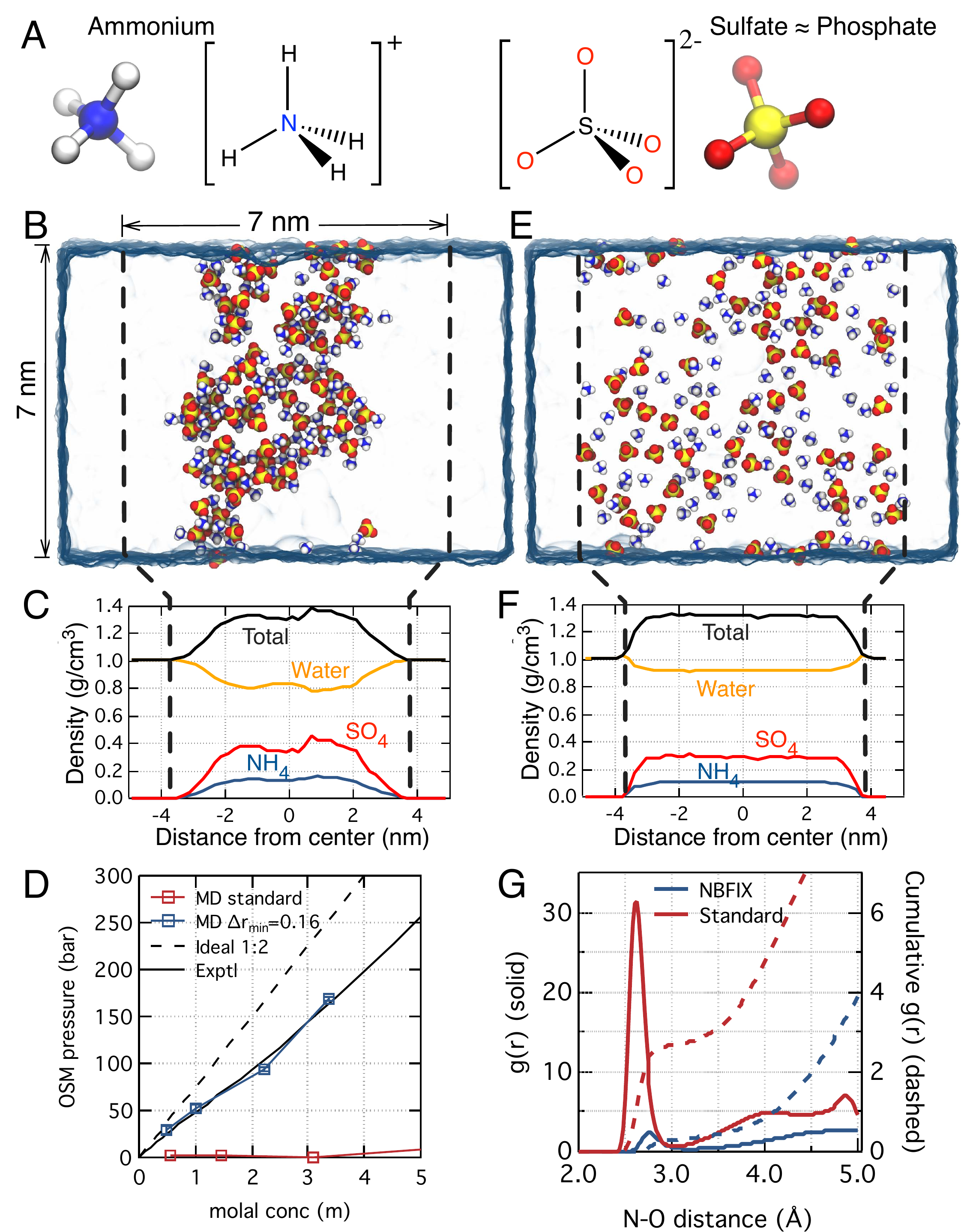}
    \caption{\label{fig:ammsul} 
     Calibration simulations of the CHARMM force field using ammonium sulphate solutions.
    (A)  Chemical structure of dissociated ammonium sulphate,  \ce{(NH4)2SO4}. 
		The oxygen and nitrogen atoms subject to the NBFIX correction are shown in red and blue, respectively.
	      The NBFIX values are given in \tab{nbfix}.
    (B) A simulation system used for calibration simulations. The simulation unit cell contains water (blue semi-transparent surface) 
    divided into two compartments by two planar half-harmonic potentials (depicted by dashed lines).
 One of the compartments contains 
  a  $\sim3$~m aqueous solution of   \ce{(NH4)2SO4} (shown as vdW spheres), the other only pure water.  The half-harmonic potentials apply restraining forces to the atoms of ammonium and sulphate molecules confining them to the same compartment;  the potentials do not act on water molecules. The force exerted by the potentials on the ammonium and sulfate molecules reports on the osmotic pressure in the
  \ce{(NH4)2SO4}-filled compartment. The snapshot illustrates the system's configuration observed
    at the end of a $\sim$25-ns simulation performed using  the standard CHARMM force field.
    For clarity, only a 10~\AA-wide slice of the system is shown. 
    (C) Density profiles of molecular species in the direction normal to the planes of the confining potentials for the system simulated using the standard CHARMM force field. Each data point represents an average over a $\sim$1.5-\AA-slice of the system and a 25-ns MD trajectory.
    (D) Osmotic pressure of a \ce{(NH4)2SO4} solution as a function of the \ce{(NH4)2SO4} concentration  obtained from MD simulations performed with (blue) and without (red) our NBFIX corrections.  For comparison, experimentally determined~\cite{ROBI1959} and ideal solution dependences of the osmotic pressure
    on concentration are shown as solid and dashed black lines, respectively.
    (E,F) Same as in panels B and C but for the simulations performed applying our NBFIX correction.
    (G) Radial distribution functions, $g(r)$ (solid lines), and cumulative $g(r)$ (dashed lines) for a $\sim$3~m \ce{(NH4)2SO4} solution simulated with (blue) and without (red) our NBFIX correction. 
%
    %
    }
\end{figure}

Calibration of the amine--phosphate interaction using a phosphate salt dissolved in water is challenging because  the phosphate group can adopt multiple protonation states  at physiological pH. For example,  aqueous solution of ammonium phosphate, \ce{(NH4)3PO4}, contains  several molecular species  (e.g., \ce{PO4^{3-}}, \ce{HPO4^{2-}}, \ce{H2PO4^{-}}, and \ce{H3PO4}) at significant concentrations, all of which  
 can simultaneously interact with the ammonium ion.  
In contrast,  aqueous solution of ammonium sulphate, \ce{(NH4)2SO4}, contains predominantly sulfate anions, \ce{SO4^{2-}}, 
and ammonium cations, \ce{NH4^{+}}, whereas the concentration of  bisulfate anions, \ce{HSO4^{-}},  is $\sim10^5$  times smaller at pH~7.
 From a structural standpoint,  the tetrahedral geometry of the four sulfate's oxygens is similar to that of the four phosphate's oxygens~\cite{CLEL2006}.
In fact, a sulfate ion is frequently found occupying  a phosphate ion's binding pocket in protein structures~\cite{LUBB2007,COPL1994,RIE2007,KUMA2010,SWAP2013}.
Furthermore,  parameters describing vdW interactions
 of oxygen atoms in carboxylate, sulfate, and phosphate groups are identical in both standard CHARMM and AMBER force fields.
Here,  we use  an aqueous solution of ammonium sulfate as a proxy for validation and  calibration of amine--phosphate interaction within the CHARMM force field.  An alternative strategy for parameterization of amine--phosphate interactions using a DNA array system is described in the next section.     


%

To validate and improve the amine--sulfate (and amine--phosphate) interactions within the CHARMM force field,  we applied the  simulation procedures described in the previous section  to ammonium sulphate solution.
 The osmotic pressure of ammonium sulphate solution was determined using the two-compartment system,~\fig{ammsul}B.
 As in the case of glycine solution, considerable aggregation effects were observed  in the simulation of a 3~m ammonium sulphate solution performed using the standard CHARMM force field,~\fig{ammsul}B,~C.
The simulated osmotic pressure showed dramatic inconsistencies with experiment~\cite{ROBI1959}  at all concentrations tested,~\fig{ammsul}D.
To remedy the problem,  we modified the LJ $\rmin$ parameter describing the vdW interaction of the  ammonium nitrogen--sulfate oxygen pair, $\rminNOS$. The experimental osmotic pressure was recovered for a $\sim$3~m solution when $\rminNOS$ was increased by 0.16~\AA~from its standard value,~Fig.~S1B. Using this new parameterization of $\rminNOS$,  we could reproduce the osmotic pressure  of ammonium sulphate solution in the full range of concentrations,~\fig{ammsul}D. The systems simulated using our NBFIX correction had homogeneous distribution of solutes in the solute compartment,~\fig{ammsul}E,F. The probability of direct contact between the ammonium nitrogen and sulphate oxygen decreased by more than an order of magnitude upon application of the NBFIX correction,~\fig{ammsul}G.



\subsection{Calibration of amine--phosphate interaction for  the AMBER force field  through simulations of  DNA--DNA interactions}

\begin{figure}[t!]
    \center
    \includegraphics[width=6.25in]{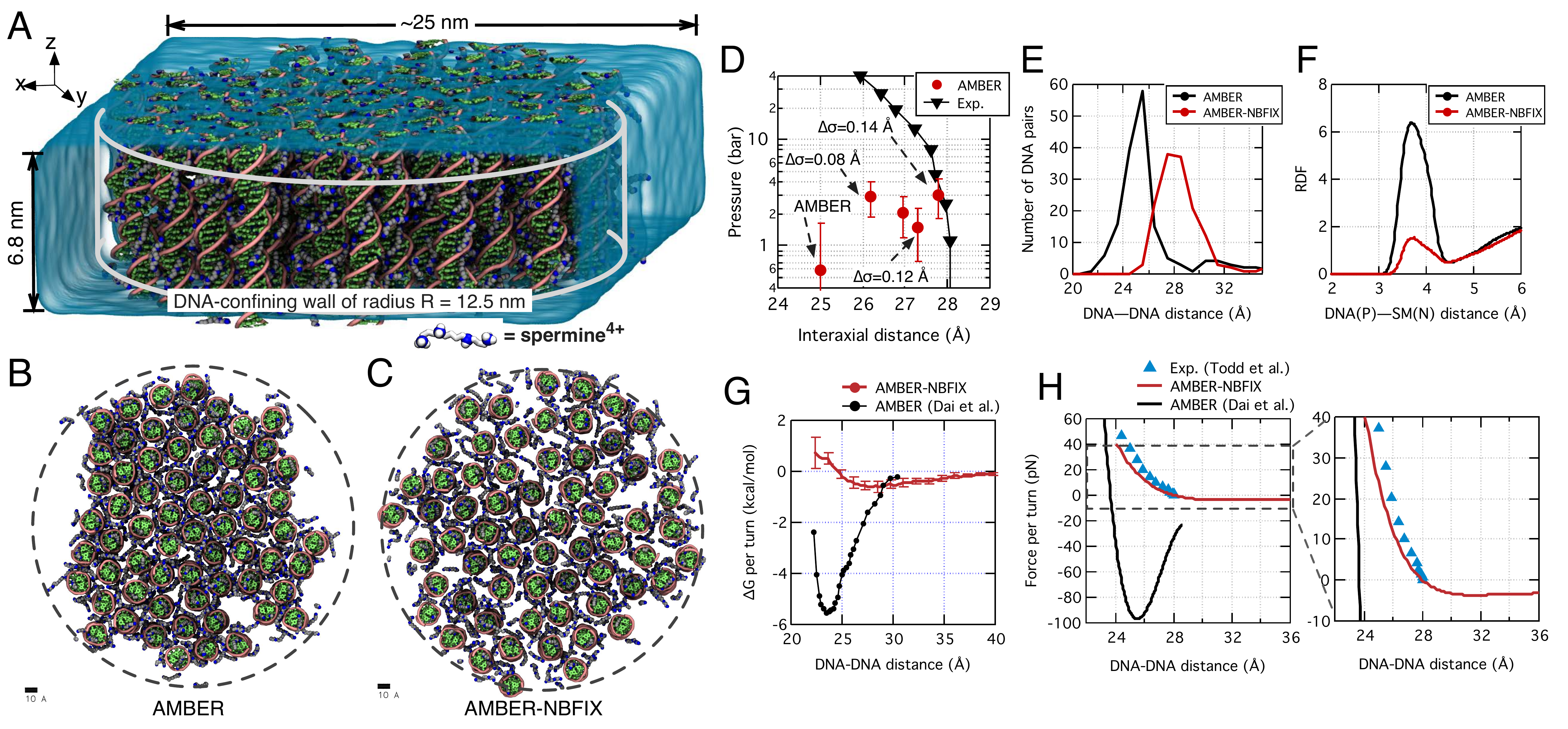}
    \caption{\label{fig:array} 
     Calibration simulations of the AMBER force field using a DNA array system.
     (A)  The simulation system containing an array of 64~dsDNA molecules (pink and green) submerged in a rectangular volume of water (semi-transparent surface). Each DNA strand is  covalent linked to itself across the periodic boundary, making the DNA molecules effectively infinite. 
The phosphorous atoms of  DNA are subject to a half-harmonic potential confining the DNA molecules to a 12.5-nm radius cylinder;  water and ions can freely pass in and out of the DNA array.
     The force applied by the half-harmonic potential to DNA reports on the internal pressure of the DNA array. 
     (B,C) Typical configurations of the DNA array observed after a 60~ns MD simulation performed using the AMBER99bsc0 force field  with ($\Delta\sigNOP = 0.14$~\AA, panel~C) and without  ($\Delta\sigNOP = 0$~\AA, panel~B) our NBFIX correction. 
     (D)  The average DNA array pressure versus the average inter-axial distance  for the specified corrections to the vdW parameter describing the interaction of spermine nitrogen with DNA phosphate oxygen, $\Delta \sigNOP$. The  experimental data (black) are taken from Ref.~\cite{RAU1984}. Using  $\Delta \sigNOP$ of 0.14~\AA~gives the best agreement   with the experimental data.
     (E) The distribution of the inter-DNA distances in the DNA array simulations performed with (red, $\Delta\sigNOP = 0.14$~\AA) and without (black) our NBFIX correction. The data were collected over the last 30~ns segments of the respective trajectories.
     (F) RDF of spermine nitrogen atoms with respect to DNA phosphorus atoms
      in the simulations of the DNA array  with (red, $\Delta\sigNOP = 0.14$~\AA) and without (black)  our NBFIX correction. 
     The RDFs were computed using the last 30~ns segments of the respective trajectories.
     (G) Potential of mean forces (PMFs) as a function of the inter-axial distance for a pair of dsDNA (dG$_{20}\cdot$dC$_{20}$) computed using the umbrella sampling method (red). For each window (1-\AA~spacing), $\sim$40-ns simulation was performed. Error bars indicate standard deviations  among four PMF curves  computed using independent 10~ns fragments of the sampling trajectories. 
     For comparison, the black line shows a PMF computed using the standard AMBER99 force field~\cite{DAI2008}.
     (H) The average inter-DNA force as a function of the inter-DNA distance. To compute the forces, the PMF curves in panel G were fitted by a double-exponential function;  the forces were obtained as numerical derivatives of the fitted functions. For comparison, experimentally measured forces at 2-mM {\SM} are shown as blue triangles~\cite{TODD2008}. 
    }
\end{figure}

In the absence of AMBER-compatible parameters describing solutions of ammonium sulphate, we performed calibration of the amine--phosphate interactions using a DNA array system, for which the experimental dependence of the osmotic pressure on the inter-DNA distance is experimentally known~\cite{TODD2008}.
Following our previous work~\cite{YOO2012}, we built a simulation system containing 64  20-bp dsDNA molecules, the neutralizing amount (640 molecules) of {\SM} and water,~\fig{array}A.
All DNA molecules were restrained to remain within a 12.5~nm radius cylinder while water and ions could freely pass in and out of the array.
 The internal pressure of the DNA array was measured by monitoring the restraining forces applied to DNA. During our MD simulations of the DNA array system, the majority of the \SM~molecules remained in proximity of DNA.  Therefore we estimate the effective concentration of \SM~in our simulations  to be in the sub-mM regime.

Experimentally, it is known that DNA molecules condense into a hexagonal array of the 28-\AA\ average inter-DNA distance at $\sim$2-mM  \SM\ concentration~\cite{TODD2008}. 
In the simulations performed using the standard AMBER99bsc0 force field, we observed a much stronger condensation,~\fig{array}B, resulting in the mean inter-DNA distance of $\sim$25~\AA~and the DNA array pressure close to zero,~\fig{array}D.
This result is consistent with the previous  MD study that found the most probable distance between two DNA molecules in \SM\ solution 
to be  $\sim$24-\AA~\cite{DAI2008}; the latter study used  AMBER99.

To determine if our NBFIX correction for the amine nitrogen--carboxylate oxygen  interaction ($\Delta\sigNOC=0.08$~\AA) could improve the agreement between simulation and experiment, we  repeated our  simulation of the DNA array using the NBFIX correction for the amine nitrogen--phosphate oxygen interaction, $\Delta\sigNOP =0.08$~\AA.
Although the  inter-DNA distance increased to $\sim$26~\AA, it was still 2-\AA~smaller than the experimental value,~\fig{array}D.
This result implies  that the NBFIX correction  for the amine--phosphate pair can differ  from that for the amine--carboxylate pair.
Thus, we gradually increased the $\sigNOP$ parameter by 0.02~\AA~from 0.08 to 0.14~\AA, performed a 60-ns MD simulation for each value of the parameter.
The average inter-DNA distance gradually increased with $\sigNOP$.
The best agreement with experiment was achieved at $\Delta \sigNOP = 0.14$~\AA,~\fig{array}D.
%
Plots of the radial distribution functions indicate a   3-fold decrease in the number of direct contacts between spermine amine and DNA phosphate groups in the simulations performed using the NBFIX correction,~\fig{array}F.

To characterize the interactions between  DNA molecules more quantitatively, we performed umbrella sampling simulations of two DNA molecules in \SM\ solution using our NBFIX correction ($\sigNOP = 0.14$~\AA), which yielded the inter-DNA potential of mean force (PMF), ~\fig{array}G. 
In these simulations,  a pair of effectively infinite dsDNA molecules (\ce{dG_{20}}$\cdot$\ce{dC_{20}}) was neutralized by ten \SM~cations.
The inter-DNA distance, $\xi$, was used as a reaction coordinate; the umbrella sampling simulations were performed for the
 22-to-40~\AA~range of inter-DNA distances with 1~\AA\ window spacing.
A similar PMF was previously computed by Dai and co-workers using the standard AMBER99 force field.~\cite{DAI2008}.
\fig{array}G shows the comparison of the two PMF's. 
The PMF obtained with and without the NBFIX correction has the free energy minima of $-0.5$ and $-5.5$~kcal/mol/turn at $\xi = 28$ and 24~\AA, respectively.

Taking numerical derivatives of the PMF's, we obtained the dependence of the effective DNA--DNA force on the inter-DNA distance.
When compared to the experimental estimates of the DNA--DNA force, the use of the NBFIX correction  is seen to dramatic improve
 accuracy of MD simulations,~\fig{array}H. 
Below the experimentally determined equilibrium inter-DNA distance ($\xi < 28$~\AA), the simulations employing the NBFIX correction  and the experiment suggest repulsive forces of comparable magnitudes,~\fig{array}H.
Conversely, the inter-DNA force computed using the standard AMBER99 force field without the NBFIX corrections predicts attractive force that can be as large as $-100$~pN at $\xi = 25$~\AA,~\fig{array}H.

\subsection{The effect of NBFIX corrections on peptide-mediated DNA--DNA interactions}

\begin{figure}[h!]
    \center
    \includegraphics[width=6.25in]{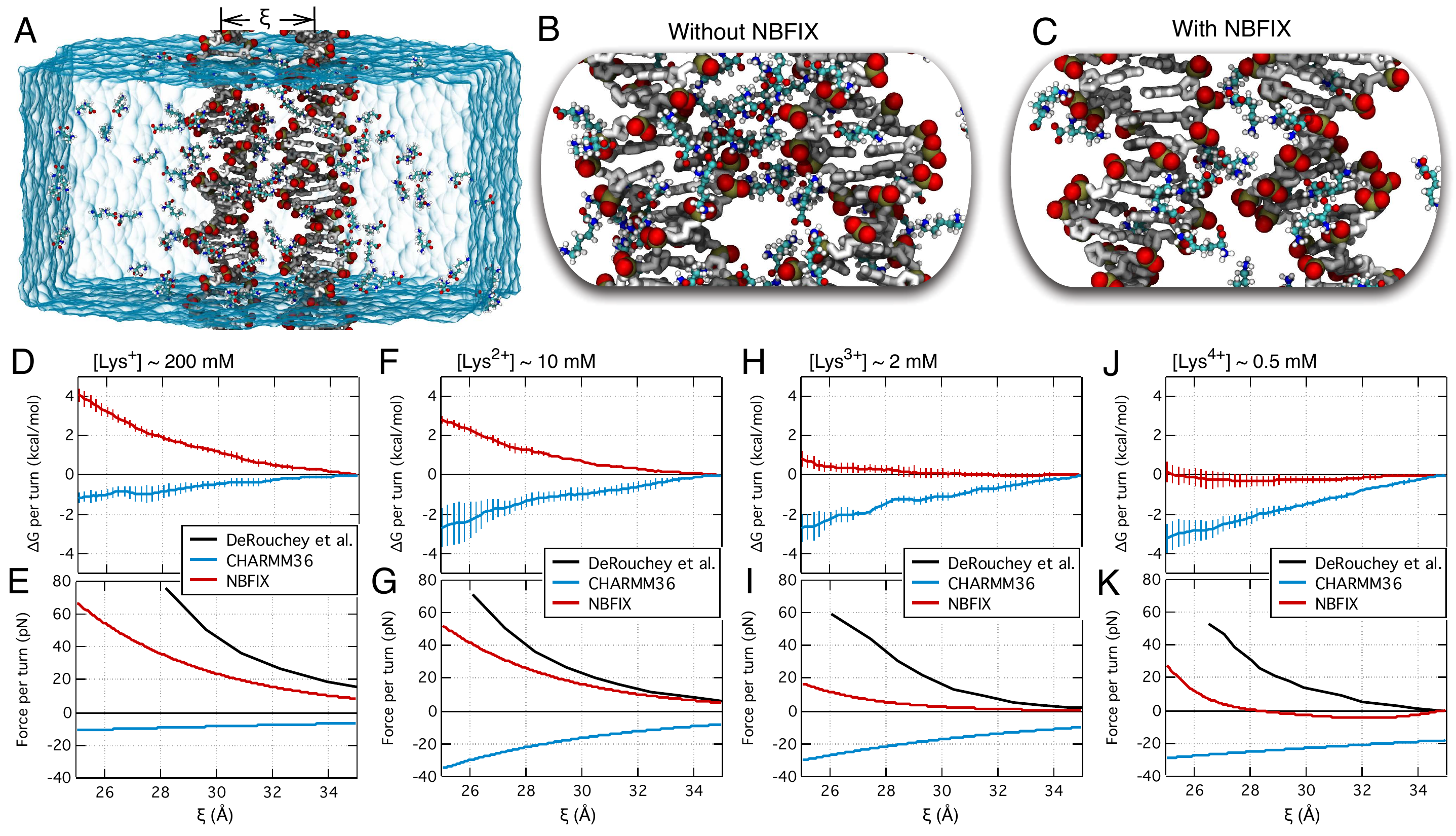}
    \caption{\label{fig:dnapmf} 
    The effects of NBFIX corrections on DNA--DNA interaction.
    (A) A typical system used for characterization of the DNA--DNA interactions. 
    A pair of dsDNA molecules  (gray)  is placed in a hexagonal volume of water (semi-transparent surface) containing 
   a 200~mM  solution of lysine monomers. The phosphorus and phosphate oxygen atoms of DNA are shown using tan and red spheres, respectively.    Lysine monomers are shown in a ball-and-stick representation; oxygen and nitrogen atoms are highlighted using red and blue spheres, respectively. A harmonic potential (not shown) maintains the distance between the helical axes of the dsDNA molecules at a specified value $\xi$. 
    (B,C) Close up views of the systems at the end of the simulations performed using the standard CHARMM36 force field (B) and with  our NBFIX corrections (C), Table~1.  In both simulations, the distance between the DNA molecules
     $\xi=28$~\AA; the concentration of lysine monomers was 200~mM. For clarity,  only those lysine monomers whose  C$\alpha$ atoms   located within 8~\AA~from DNA  are shown.
    (D--K) The interaction free energy, $\Delta G$, (D,F,H and J) and the effective DNA--DNA force, $F$, (E,G,I and K) between two dsDNA molecules as a function of the inter-DNA distance, $\xi$.
    The $\Delta G$ and $F$ values computed from the simulations with and without our NBFIX correction are shown as red and blue lines, respectively.
    The  experimental DNA--DNA force (black lines in E,G,I and K) were taken from Ref.~\cite{DERO2013}.
     The electrolyte conditions in panels D and E, F and G, H and I, and J and K were [Lys$^{+}$] $\sim$ 200~mM, [Lys$^{2+}$] $\sim$ 10~mM, [Lys$^{3+}$] $\sim$ 2~mM and  [Lys$^{4+}$] $\sim$ 0.5~mM, respectively.
    The interaction free energy was computed from a set of umbrella sampling simulations; the duration of the MD trajectory in each umbrella sampling window  was 40~ns.
    The error bars in the $\Delta G$ curves indicate the standard deviations among four $\Delta G$ estimates obtained using independent 10~ns fragments of the umbrella sampling trajectories.
    To compute the DNA--DNA forces, the corresponding interaction free energy  was fitted to a double exponential function~\cite{RAU1984}; numerical differentiation of the fitting function with respect to $\xi$ yielded the effective force as $F = -dG/d\xi$.
    }
\end{figure}

Examples of biomolecular systems that permit direct comparison of  quantitative experimental information  to the outcome of a microscopic simulation remain scarce.  
Thus, the association free energy of two biomacromolecules is readily available from experiment but it is cumbersome and expensive to determine computationally, although such calculations are becoming increasingly common~\cite{MAFF2012A}. On the other hand,  precise information about the forces between biomacromolecules arranged in a specific conformation are readily available from MD simulations but are difficult to obtain  experimentally.

Here, we use experimental DNA array data~\cite{DERO2013} to assess the improvements in the description of DNA--protein systems brought about by our set of NBFIX corrections.  The pairwise forces between double stranded DNA (dsDNA) sensitively depend on the solution environment and the distance between the DNA molecules~\cite{RAU1984,TODD2008,LUAN2008C,MAFF2010B}. The interactions are repulsive in monovalent salt, but can turn attractive in the present of polycations~\cite{RAU1984}. One class of such polycations are lysine oligomers. Experimentally, it has been determined that interactions between dsDNA in a solution of lysine monomers or lysine dimers is repulsive, but it turns attractive as the length of lysine peptides increases~\cite{DERO2013}. 
For example,  DNA condensation in 2~mM tri-lysine solution is marginal and is characterized by the equilibrium inter-DNA distance of 39~\AA.  At the same time, DNA condensation in 0.1~mM hexa-lysine is relatively stable and is characterized by the equilibrium inter-DNA distance of 32~\AA~\cite{DERO2013}.

To compute the interaction free energy of two dsDNA molecules, $\Delta G^\mathrm{DNA-DNA}$,  we constructed a simulation
system containing a pair of 21-bp dsDNA molecules arranged parallel to each other and made effectively infinite by the covalent bonds across the periodic boundary of the system. The volume of the simulation cell was filled with electrolyte of desired composition and concentration,~\fig{dnapmf}A.
The potential of mean force (PMF) 
as a function of the DNA--DNA distance  was determined through a set of umbrella sampling simulations, see Methods for details.
In the case of a 200~mM [Lys$^{+}$] solution, the difference between the simulations performed using the standard CHARMM36 force field
with and without our NBFIX corrections can be discerned by visual inspection.    
The simulations performed without  the NBFIX corrections are characterized by significant aggregation of lysine monomers at the surface of DNA,~\fig{dnapmf}B and Fig.~S2A, as well as self-aggregation of lysine monomers,~\fig{dnapmf}B and Fig.~S2B.
Using the NBFIX corrections considerably reduces  the aggregation propensity of lysine monomers,~\fig{dnapmf}C and Fig.~S2A,B.

The calculations of the inter-DNA PMF show qualitative differences in the behavior of the systems simulated with and without the NBFIX corrections. At 200 mM [Lys$^{+}$],  the PMF obtained using standard CHARMM decreases monotonically with the inter-DNA distance,~\fig{dnapmf}D, indicating that spontaneous association of DNA molecules is energetically favorable. An opposite behavior is observed when NBFIX corrections are used: the PMF monotonically increases as the DNA--DNA distance deceases, indicating mutual repulsion of the molecules. Experimentally, spontaneous association of DNA molecules was not observed in the presence of lysine monomers~\cite{DERO2013}. Thus, the standard parameterization of the CHARMM force field predicts simulation outcomes that are in 
qualitative disagreement with experiment.   

To quantitatively compare the simulation results to experiment,  we computed the dependence of the DNA--DNA effective force, $F^\mathrm{DNA-DNA}$, on the DNA--DNA distance by numerically differentiating the DNA--DNA PMF. 
Whereas the experimental DNA--DNA force is repulsive ($> 0$) and increases as the distance between DNA molecules becomes smaller~\cite{DERO2013}, 
$F^\mathrm{DNA-DNA}$ computed using the standard CHARMM36 is attractive and approximately constant ($-10$~pN per turn),~\fig{dnapmf}E.
The force-versus-distance dependence obtained using NBFIX is in a much better agreement with experiment: the force is repulsive and decreases with the DNA--DNA distance,~\fig{dnapmf}E. 
Even with the NBFIX corrections enabled, however, the simulated and experimentally measured forces remain quantitatively different:  the simulated forces underestimate the DNA--DNA repulsion by a factor of two. 
Possibly, this result indicates that parameters describing  non-bonded or bonded interactions of lysine side chains still have some room for improvement.
Similar results were obtained from the simulations of a DNA pair at 10~mM [Lys$^{2+}$] (\fig{dnapmf}F,G), 2~mM [Lys$^{3+}$] (\fig{dnapmf}H,I), and 0.5~mM [Lys$^{4+}$] (\fig{dnapmf}J,K): the NBFIX corrections clearly brought the results of the simulations closer to experimental reality. 

The significant discrepancy of the standard CHARMM36 force field  in describing  the lysine-mediated DNA--DNA forces 
results from the excessive direct pairing of lysine monomers and DNA phosphate groups,~\fig{dnapmf}B and  Fig.~S2A.
Similar behavior was previously reported in MD simulations of DNA arrays in the presence of monovalent and divalent cations~\cite{YOO2012,YOO2015A}.
Excessive direct contacts are also responsible for the self-association behavior of lysine monomers observed in the simulations using 
standard CHARMM: the lysine monomers cluster because of the artificially strong amine--carboxylate interactions,~Fig.~S2B. 
The application of  the NBFIX corrections significantly reduces the amount of  direct  paring between lysine amine and DNA phosphate and between lysine amine and lysine carboxylate,~\fig{dnapmf}C and  Fig.~S2A.


Repeating  the  simulations of the 10~mM [Lys$^{2+}$] and 2~mM [Lys$^{3+}$] systems using the AMBER99 force field produced
similar outcomes as the simulations performed using the CHRAMM force field. 
For both peptides, the AMBER99 force field showed significant inter-DNA attractions, which contradicts experimental observations,~Fig.~S3A,B,E,F. The artificial attraction was caused by the overestimated propensity for direct pair formation between lysine amine and DNA phosphate groups, Fig.~S3C,G, and between the peptides,~Fig.~S3D,H.
Applying our NBFIX correction significantly improved the agreement between simulation and experiment, leaving, however, some room for further refinement,~Fig.~S3B,F.

The residual quantitative discrepancy between the simulations employing  our NBFIX corrections and experiment suggests that other  interactions may also require refinement, for example, non-bonded interactions between nonpolar carbon atoms or bonded interactions within the peptides. 
For AMBER force fields, it has been known that uncharged peptides tend to form artificial aggregates~\cite{NERE2012,BEST2014}.
Best et al. tried to remedy the problem by scaling LJ $\epsilon$ parameters for all peptide atoms and water oxygen pairs by the same factor (1.10) to reproduce the experimentally measured radius of gyration of a 34-residue-long peptide~\cite{BEST2014}.
In contrast, our approach to force field refinement has the advantage of targeting specific interactions.
We have singled out amine--carboxylate and amine--phosphate interactions using the simplest possible model systems, which allowed us to target these specific interactions for refinement.
It is very possible that similar corrections can be developed for other types of specific interaction and mixed without interfering with the existing corrections. 
In this regard, experimental measurements of osmotic pressure in simple polymer solutions can provide the  experimental data
required for specific refinement of the nonpolar carbon interactions.

\subsection{The effect of NBFIX corrections on MD simulations of lipid bilayer membranes}

\begin{figure}[t!]
    \center
    \includegraphics[width=6.5in]{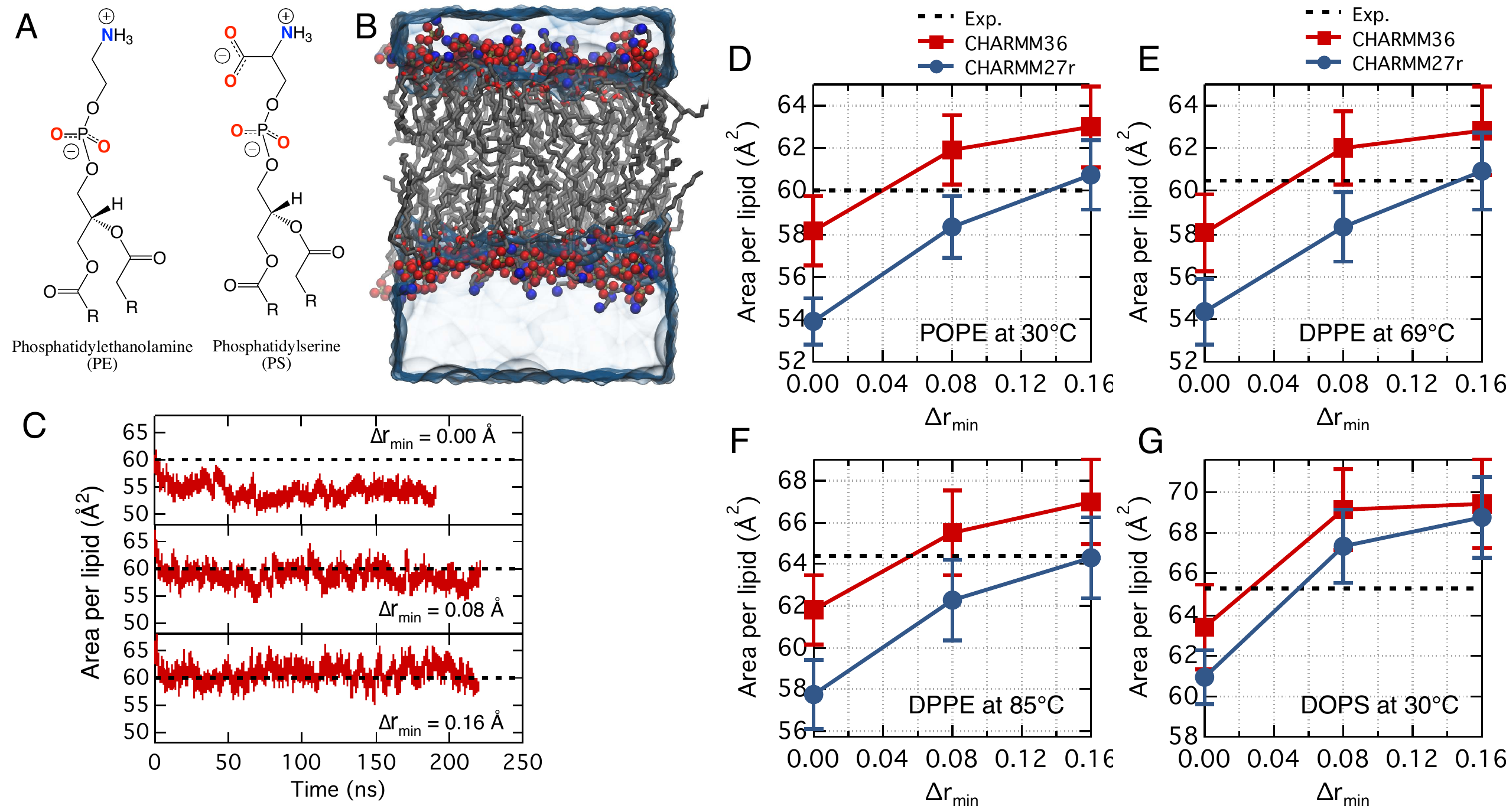}
    \caption{\label{fig:lipid} 
    The effects of NBFIX corrections on MD simulations of lipid bilayers membranes. 
    (A) The chemical structures of neutral phosphatidylethanolamine (PE) and anionic phosphatidylserine (PS) lipid head groups.
The amine nitrogen and phosphate oxygen atoms subject to the NBFIX corrections are highlighted in blue and red, respectively; R indicates a hydrocarbon tail.
 (B)   A representative conformation of a DPPE bilayer.
 DPPE molecules are shown in gray molecular bonds with phosphate oxygen and amine nitrogen atoms highlighted as red and blue spheres, respectively. The unit cell's water box is shown as a semi-transparent surface.
  (C) Area per lipid as a function of simulation time for several parameterizations of the amine nitrogen--phosphate oxygen  interactions,  $\Delta \rminNOP$. 
  The simulations were performed using the CHARMM27r parameter set~\cite{KLAU2005} at 30\celsius\
  and the following three   values of $\Delta \rminNOP$:  0.00 (no NBFIX correction), 0.08, and 0.16~\AA. 
Dashed lines indicate  the area per lipid  measured experimentally~\cite{KLAU2010}.
    (D--G) The average area per lipid  as a function of  $\Delta \rminNOP$ 
    in MD simulations of the following lipid membranes:   POPE at 30\celsius\ (D), DPPE at 69\celsius\ (E), DPPE at 85\celsius\ (F), and DOPS  at 30~\celsius\ (G).
    Note that $\Delta \rminNOC$ = 0.08~\AA~was applied only when $\Delta \rminNOP$ = 0.08 or 0.16~\AA.
    The simulation data obtained using the CHARMM27r~\cite{KLAU2005} and CHARMM36~\cite{KLAU2010} lipid parameter sets are shown in blue and red, respectively. The error bars indicate the standard deviation of the 1-ns block averages of the 2-ps sampled area-per-lipid data.
    Experimental data for POPE, DPPE, and DOPS were taken from Ref.~\cite{KLAU2010}, Ref.~\cite{PETR2000}, and Ref.~\cite{PETR2004}, respectively.
    }
\end{figure}

In this section, we describe the effects of our NBFIX corrections on the structure of simulated lipid bilayer membranes containing amine and/or carboxylate groups. \fig{lipid}A illustrates the two lipid head groups  considered: neutral phosphoethanolamine (PE) and anionic phosphatidylserine (PS). In total, four different lipid membranes were simulated:  two  PE lipids, 1,2-dipalmitoyl-sn-glycero-3-phosphoethanolamine (DPPE) and 1-palmitoyl-2-oleoyl-sn-glycero-3-phosphoethanolamine (POPE), and two  PS  lipids, 1,2-dioleoyl-sn-glycero-3-phosphoserine (DOPS) and 
1-palmitoyl-2-oleoyl-sn-glycero-3-phosphoserine (POPS). 

\fig{lipid}B illustrates a simulation system containing 72 DPPE molecules. 
For rigorous comparison with the existing variants of the CHARMM lipid force field, all simulations described in this section were
performed using the CHARMM package~\cite{BROO2009b} and the simulation setup exactly identical to that used for the development of the CHARMM lipid force field by Klauda and co-workers~\cite{KLAU2010}. In addition to CHARMM36, we also tested the CHARMM27r variant of the force field specifically optimized for simulations of  lipid bilayers~\cite{KLAU2005}. The Methods section provides a more detailed description of the simulation setup. 

To quantify the quality of the force field and the effect of our NBFIX correction, we monitored the lipid density (or area per lipid), which is, historically, the most important target in the development of lipid force fields~\cite{KLAU2008}. \fig{lipid}C plots the time dependence of the area per lipid for a POPE bilayer simulated using the CHARMM27r lipid force field and $\Delta \rminNOP = 0.00, 0.08$, and 0.16~\AA~for the amine nitrogen--phosphate oxygen pair. Clearly, the area per lipid increased with $\Delta \rminNOP$, indicating reduced attraction between the lipid head groups. Remarkably,  at $\Delta\rminNOP = 0.16$~\AA, the computed area per lipid matches the experimental value (\fig{lipid}C, dashed lines). Similar increases of the area per lipid with  $\rminNOP$ were observed  in the simulations of the other three lipid bilayer systems, Figs.~S3, S4 and S5.

Panels D--F of \fig{lipid} plot the  average area per lipid for POPE and DPPE bilayers as a function of $\Delta \rminNOP$ computed using the CHARMM27r and CHARMM36 force fields. In all cases, the area per lipid  monotonically increases with $\Delta \rminNOP$. Regardless of the lipid type (POPE or DPPE) and temperature, CHARMM27r results match the experimental data (dashed lines) best with $\Delta \rminNOP = 0.16$~\AA, suggesting that the NBFIX correction of 0.16~\AA~obtained from the simulations of ammonium sulphate solutions 
can also apply to the simulations of lipid bilayer membranes. The area per lipid computed using the CHARMM36 force field was
$2-4$~\AA$^2$ larger than that obtained using the CHARMM27r force field, which was expected because both bonded and non-bonded parameters near carbonyl groups were modified in CHARMM36 to increase the area per lipid values~\cite{KLAU2010}. 
Note that amine--carboxylate and amine--phosphate interactions are identical in both CHARMM27r and CHARMM36.
Comparison of the computed and experimental deuterium order parameters of lipid tails also shows best agreement for the CHARMM27r force field with $\Delta \rminNOP = 0.16$~\AA, Fig.~S6A,B.

The interactions within the anionic PS lipid bilayer membranes are intrinsically more complex than those of PE lipids because each PS lipid molecule has an additional carboxylate group.
Similar to the PE lipids, the area per lipid values of the DOPS and POPS bilayers increase monotonically as  $\Delta \rminNOP$ increases,~\fig{lipid}G and Fig.~S7A,B. 
Note that, in our simulations of DOPS and POPS bilayers, the $\Delta \rminNOC = 0.08$~\AA~correction for the amine--carboxylate interaction was applied simultaneously with the $\Delta \rminNOP$ = 0.08 or 0.16~\AA~correction for amine--phosphate interaction.
For the DOPS bilayer  simulated with the $\Delta \rminNOP$ and $\Delta \rminNOC$ corrections, the area per lipid was larger than the experimental value by $<4$~\AA$^2$. For the POPS bilayer, the deuterium order parameter also indicates that the POPS membranes are slightly overstretched when the $\Delta \rminNOP$ and $\Delta \rminNOC$ correction is applied, Fig.~S8C--E. 

Although our study does not provide a general solution to the problems remaining in the development of the lipid force field, our results 
highlight the importance of proper parameterization of  the amine--carboxylate and amine--phosphate interactions for accurate description of  the lipid head group packing. Accurate reproduction of the local arrangement of the lipid head groups, such as provided by our NBFIX corrections, can be of importance for simulations of lipid head group recognition by proteins~\cite{MCLA2005} and for simulations of lipid mixtures, where minor differences in the interactions of the head groups can differentiate lipid mixing from phase segregation~\cite{BROW1998}.

\section {Conclusion}



In this work, we presented an improved parametrization of amine--carboxylate and amine--phosphate interactions for MD simulations of biomolecular systems. Because amine, carboxylate, and phosphate groups are essential chemical groups of all classes of biomolecules---including proteins, nucleic acids, and lipids---our refined parameters will be of use in computational studies of a broad range of biomolecular systems. In comparison to the standard models, a particularly noticeable improvement  was demonstrated in description of dense DNA systems and peptide-mediated DNA--DNA forces, potentially 
 enhancing the realist of future MD simulations of a variety  of nucleic acid systems, including transcription factors, polymerases, and motor proteins.

In addition to improving accuracy of computational models of DNA--protein systems, our NBFIX corrections can also be applied to the simulations of protein--lipid interactions where both amine--phosphate and amine--carboxylate interactions  play central roles. For lipid--lipid interactions, our results demonstrates that application of our NBFIX corrections to the CHARMM27r force field  certainly improves the accuracy of the latter.
For CHARMM36 lipid force field, further investigations will be required to determine how  our improved description of amine--phosphate interactions  can be incorporated within the existing model.
Overall, the results of our study strongly suggest that revised parameterization of  amine--carboxylate and amine--phosphate interactions should be considered in the future development of the CHARMM and AMBER force fields.


\begin{acknowledgement}
This work was supported by the National Science Foundation grant PHY-1430124. The authors acknowledge supercomputer time at the Blue Waters  Sustained Petascale Facility (University of Illinois) and 
at the Texas Advanced Computing Center (Stampede, allocation award MCA05S028).
We thank Dr. Jeffery Klauda for sharing his input file for the CHARMM simulations.
\end{acknowledgement}

\begin{suppinfo}
Detailed description of the setup and procedures used in our MD simulations,  discussion of the force field choices and our treatment of magnesium, complete account of the fitting procedures, and sample parameter files with NBFIX corrections. This material is available free of charge via the Internet at http://pubs.acs.org.
\end{suppinfo}

\providecommand{\latin}[1]{#1}
\providecommand*\mcitethebibliography{\thebibliography}
\csname @ifundefined\endcsname{endmcitethebibliography}
  {\let\endmcitethebibliography\endthebibliography}{}


\begin{mcitethebibliography}{92}
\providecommand*\natexlab[1]{#1}
\providecommand*\mciteSetBstSublistMode[1]{}
\providecommand*\mciteSetBstMaxWidthForm[2]{}
\providecommand*\mciteBstWouldAddEndPuncttrue
  {\def\EndOfBibitem{\unskip.}}
\providecommand*\mciteBstWouldAddEndPunctfalse
  {\let\EndOfBibitem\relax}
\providecommand*\mciteSetBstMidEndSepPunct[3]{}
\providecommand*\mciteSetBstSublistLabelBeginEnd[3]{}
\providecommand*\EndOfBibitem{}
\mciteSetBstSublistMode{f}
\mciteSetBstMaxWidthForm{subitem}{(\alph{mcitesubitemcount})}
\mciteSetBstSublistLabelBeginEnd
  {\mcitemaxwidthsubitemform\space}
  {\relax}
  {\relax}

\bibitem[Pabo and Sauer(1984)Pabo, and Sauer]{PABO1984}
Pabo,~C.~O.; Sauer,~R.~T. \emph{Annu.\ Rev.\ Biochem} \textbf{1984}, \emph{53},
  293--321\relax
\mciteBstWouldAddEndPuncttrue
\mciteSetBstMidEndSepPunct{\mcitedefaultmidpunct}
{\mcitedefaultendpunct}{\mcitedefaultseppunct}\relax
\EndOfBibitem
\bibitem[Pabo and Sauer(1992)Pabo, and Sauer]{PABO1992}
Pabo,~C.~O.; Sauer,~R.~T. \emph{Annu.\ Rev.\ Biochem} \textbf{1992}, \emph{61},
  1053--1095\relax
\mciteBstWouldAddEndPuncttrue
\mciteSetBstMidEndSepPunct{\mcitedefaultmidpunct}
{\mcitedefaultendpunct}{\mcitedefaultseppunct}\relax
\EndOfBibitem
\bibitem[Thomsen and Berger(2009)Thomsen, and Berger]{THOM2009}
Thomsen,~N.~D.; Berger,~J.~M. \emph{Cell} \textbf{2009}, \emph{139},
  523--34\relax
\mciteBstWouldAddEndPuncttrue
\mciteSetBstMidEndSepPunct{\mcitedefaultmidpunct}
{\mcitedefaultendpunct}{\mcitedefaultseppunct}\relax
\EndOfBibitem
\bibitem[Luger \latin{et~al.}(1997)Luger, Mader, Richmond, Sargent, and
  Richmond]{LUGE1997}
Luger,~K.; Mader,~A.~W.; Richmond,~R.~K.; Sargent,~D.~F.; Richmond,~T.~J.
  \emph{Nature} \textbf{1997}, \emph{389}, 251--260\relax
\mciteBstWouldAddEndPuncttrue
\mciteSetBstMidEndSepPunct{\mcitedefaultmidpunct}
{\mcitedefaultendpunct}{\mcitedefaultseppunct}\relax
\EndOfBibitem
\bibitem[Bintu \latin{et~al.}(2012)Bintu, Ishibashi, Dangkulwanich, Wu,
  Lubkowska, Kashlev, and Bustamante]{BINT2012}
Bintu,~L.; Ishibashi,~T.; Dangkulwanich,~M.; Wu,~Y.-Y.~Y.; Lubkowska,~L.;
  Kashlev,~M.; Bustamante,~C. \emph{Cell} \textbf{2012}, \emph{151},
  738--49\relax
\mciteBstWouldAddEndPuncttrue
\mciteSetBstMidEndSepPunct{\mcitedefaultmidpunct}
{\mcitedefaultendpunct}{\mcitedefaultseppunct}\relax
\EndOfBibitem
\bibitem[Rincon-Restrepo \latin{et~al.}(2011)Rincon-Restrepo, Mikhailova,
  Bayley, and Maglia]{RINC2011}
Rincon-Restrepo,~M.; Mikhailova,~E.; Bayley,~H.; Maglia,~G. \emph{Nano Lett.}
  \textbf{2011}, \emph{11}, 746--750\relax
\mciteBstWouldAddEndPuncttrue
\mciteSetBstMidEndSepPunct{\mcitedefaultmidpunct}
{\mcitedefaultendpunct}{\mcitedefaultseppunct}\relax
\EndOfBibitem
\bibitem[Bhattacharya \latin{et~al.}(2012)Bhattacharya, Derrington, Pavlenok,
  Niederweis, Gundlach, and Aksimentiev]{BHAT2012}
Bhattacharya,~S.; Derrington,~I.~M.; Pavlenok,~M.; Niederweis,~M.;
  Gundlach,~J.~H.; Aksimentiev,~A. \emph{{ACS} Nano} \textbf{2012}, \emph{6},
  6960--6968\relax
\mciteBstWouldAddEndPuncttrue
\mciteSetBstMidEndSepPunct{\mcitedefaultmidpunct}
{\mcitedefaultendpunct}{\mcitedefaultseppunct}\relax
\EndOfBibitem
\bibitem[Perutz(1978)]{PERU1978}
Perutz,~M. \emph{Science} \textbf{1978}, \emph{201}, 1187--1191\relax
\mciteBstWouldAddEndPuncttrue
\mciteSetBstMidEndSepPunct{\mcitedefaultmidpunct}
{\mcitedefaultendpunct}{\mcitedefaultseppunct}\relax
\EndOfBibitem
\bibitem[Sheinerman \latin{et~al.}(2000)Sheinerman, Norel, and Honig]{SHEI2000}
Sheinerman,~F.~B.; Norel,~R.; Honig,~B. \emph{Curr.\ Opin.\ Struct.\ Biol.}
  \textbf{2000}, \emph{10}, 153 -- 159\relax
\mciteBstWouldAddEndPuncttrue
\mciteSetBstMidEndSepPunct{\mcitedefaultmidpunct}
{\mcitedefaultendpunct}{\mcitedefaultseppunct}\relax
\EndOfBibitem
\bibitem[Petkova \latin{et~al.}(2002)Petkova, Ishii, Balbach, Antzutkin,
  Leapman, Delaglio, and Tycko]{PETK2002}
Petkova,~A.~T.; Ishii,~Y.; Balbach,~J.~J.; Antzutkin,~O.~N.; Leapman,~R.~D.;
  Delaglio,~F.; Tycko,~R. \emph{Proceedings of the National Academy of
  Sciences} \textbf{2002}, \emph{99}, 16742--16747\relax
\mciteBstWouldAddEndPuncttrue
\mciteSetBstMidEndSepPunct{\mcitedefaultmidpunct}
{\mcitedefaultendpunct}{\mcitedefaultseppunct}\relax
\EndOfBibitem
\bibitem[Ji \latin{et~al.}(1998)Ji, Grossmann, and Ji]{JI1998}
Ji,~T.~H.; Grossmann,~M.; Ji,~I. \emph{J.~Biol.\ Chem.} \textbf{1998},
  \emph{273}, 17299--17302\relax
\mciteBstWouldAddEndPuncttrue
\mciteSetBstMidEndSepPunct{\mcitedefaultmidpunct}
{\mcitedefaultendpunct}{\mcitedefaultseppunct}\relax
\EndOfBibitem
\bibitem[Bokoch \latin{et~al.}(2010)Bokoch, Zou, Rasmussen, Liu, Nygaard,
  Rosenbaum, Fung, Choi, Thian, Kobilka, Puglisi, Weis, Pardo, Prosser,
  Mueller, and Kobilka]{BOKO2010}
Bokoch,~M.~P.; Zou,~Y.; Rasmussen,~S. G.~F.; Liu,~C.~W.; Nygaard,~R.;
  Rosenbaum,~D.~M.; Fung,~J.~J.; Choi,~H.-J.~J.; Thian,~F.~S.; Kobilka,~T.~S.;
  Puglisi,~J.~D.; Weis,~W.~I.; Pardo,~L.; Prosser,~R.~S.; Mueller,~L.;
  Kobilka,~B.~K. \emph{Nature} \textbf{2010}, \emph{463}, 108--12\relax
\mciteBstWouldAddEndPuncttrue
\mciteSetBstMidEndSepPunct{\mcitedefaultmidpunct}
{\mcitedefaultendpunct}{\mcitedefaultseppunct}\relax
\EndOfBibitem
\bibitem[Kyte and Doolittle(1982)Kyte, and Doolittle]{kYTE1982}
Kyte,~J.; Doolittle,~R.~F. \emph{J.~Mol.\ Biol.} \textbf{1982}, \emph{157}, 105
  -- 132\relax
\mciteBstWouldAddEndPuncttrue
\mciteSetBstMidEndSepPunct{\mcitedefaultmidpunct}
{\mcitedefaultendpunct}{\mcitedefaultseppunct}\relax
\EndOfBibitem
\bibitem[Hessa \latin{et~al.}(2005)Hessa, Kim, Bihlmaier, Lundin, Boekel,
  Andersson, Nilsson, White, and von Heijne]{HESS2005}
Hessa,~T.; Kim,~H.; Bihlmaier,~K.; Lundin,~C.; Boekel,~J.; Andersson,~H.;
  Nilsson,~I.; White,~S.~H.; von Heijne,~G. \emph{Nature} \textbf{2005},
  \emph{433}, 377--381\relax
\mciteBstWouldAddEndPuncttrue
\mciteSetBstMidEndSepPunct{\mcitedefaultmidpunct}
{\mcitedefaultendpunct}{\mcitedefaultseppunct}\relax
\EndOfBibitem
\bibitem[von Heijne(2006)]{VONH2006}
von Heijne,~G. \emph{Nat.\ Rev.\ Mol.\ Cell Biol.} \textbf{2006}, \emph{7},
  909--18\relax
\mciteBstWouldAddEndPuncttrue
\mciteSetBstMidEndSepPunct{\mcitedefaultmidpunct}
{\mcitedefaultendpunct}{\mcitedefaultseppunct}\relax
\EndOfBibitem
\bibitem[Schmidt \latin{et~al.}(2006)Schmidt, Jiang, and MacKinnon]{SCHM2006}
Schmidt,~D.; Jiang,~Q.-X. .~X.; MacKinnon,~R. \emph{Nature} \textbf{2006},
  \emph{444}, 775--779\relax
\mciteBstWouldAddEndPuncttrue
\mciteSetBstMidEndSepPunct{\mcitedefaultmidpunct}
{\mcitedefaultendpunct}{\mcitedefaultseppunct}\relax
\EndOfBibitem
\bibitem[McLaughlin and Murray(2005)McLaughlin, and Murray]{MCLA2005}
McLaughlin,~S.; Murray,~D. \emph{Nature} \textbf{2005}, \emph{438},
  605--11\relax
\mciteBstWouldAddEndPuncttrue
\mciteSetBstMidEndSepPunct{\mcitedefaultmidpunct}
{\mcitedefaultendpunct}{\mcitedefaultseppunct}\relax
\EndOfBibitem
\bibitem[Freites \latin{et~al.}(2005)Freites, Tobias, {von Heijne}, and
  White]{FREI2005}
Freites,~J.~A.; Tobias,~D.~J.; {von Heijne},~G.; White,~S.~H. \emph{Proc.\
  Natl.\ Acad.\ Sci.\ U.S.A.} \textbf{2005}, \emph{102}, 15059--15064\relax
\mciteBstWouldAddEndPuncttrue
\mciteSetBstMidEndSepPunct{\mcitedefaultmidpunct}
{\mcitedefaultendpunct}{\mcitedefaultseppunct}\relax
\EndOfBibitem
\bibitem[Wang \latin{et~al.}(2015)Wang, Penmatsa, and Gouaux]{WANG2015}
Wang,~K.~H.; Penmatsa,~A.; Gouaux,~E. \emph{Nature} \textbf{2015}, \emph{521},
  322--7\relax
\mciteBstWouldAddEndPuncttrue
\mciteSetBstMidEndSepPunct{\mcitedefaultmidpunct}
{\mcitedefaultendpunct}{\mcitedefaultseppunct}\relax
\EndOfBibitem
\bibitem[Betz(1990)]{BETZ1990}
Betz,~H. \emph{Neuron} \textbf{1990}, \emph{5}, 383 -- 392\relax
\mciteBstWouldAddEndPuncttrue
\mciteSetBstMidEndSepPunct{\mcitedefaultmidpunct}
{\mcitedefaultendpunct}{\mcitedefaultseppunct}\relax
\EndOfBibitem
\bibitem[Armstrong \latin{et~al.}(1998)Armstrong, Sun, Chen, and
  Gouaux]{ARMS1998}
Armstrong,~N.; Sun,~Y.; Chen,~G.-Q. .~Q.; Gouaux,~E. \emph{Nature}
  \textbf{1998}, \emph{395}, 913--917\relax
\mciteBstWouldAddEndPuncttrue
\mciteSetBstMidEndSepPunct{\mcitedefaultmidpunct}
{\mcitedefaultendpunct}{\mcitedefaultseppunct}\relax
\EndOfBibitem
\bibitem[Cornell \latin{et~al.}(1995)Cornell, Cieplak, Bayly, Gould, Merz,
  Ferguson, Spellmeyer, Fox, Caldwell, and Kollman]{CORN1995}
Cornell,~W.~D.; Cieplak,~P.; Bayly,~C.~I.; Gould,~I.~R.; Merz,~K.~M.;
  Ferguson,~D.~M.; Spellmeyer,~D.~C.; Fox,~T.; Caldwell,~J.~W.; Kollman,~P.~A.
  \emph{J.~Am.\ Chem.\ Soc.} \textbf{1995}, \emph{117}, 5179--5197\relax
\mciteBstWouldAddEndPuncttrue
\mciteSetBstMidEndSepPunct{\mcitedefaultmidpunct}
{\mcitedefaultendpunct}{\mcitedefaultseppunct}\relax
\EndOfBibitem
\bibitem[Hornak \latin{et~al.}(2006)Hornak, Abel, Okur, Strockbine, Roitberg,
  and Simmerling]{HORN2006}
Hornak,~V.; Abel,~R.; Okur,~A.; Strockbine,~B.; Roitberg,~A.; Simmerling,~C.
  \emph{Proteins:\ Struct.,\ Func.,\ Bioinf.} \textbf{2006}, \emph{65},
  712--25\relax
\mciteBstWouldAddEndPuncttrue
\mciteSetBstMidEndSepPunct{\mcitedefaultmidpunct}
{\mcitedefaultendpunct}{\mcitedefaultseppunct}\relax
\EndOfBibitem
\bibitem[Buck \latin{et~al.}(2006)Buck, Bouguet-Bonnet, Pastor, and
  MacKerell]{BUCK2006}
Buck,~M.; Bouguet-Bonnet,~S.; Pastor,~R.~W.; MacKerell,~A.~D.
  \emph{Biophys.~J.} \textbf{2006}, \emph{90}, L36--38\relax
\mciteBstWouldAddEndPuncttrue
\mciteSetBstMidEndSepPunct{\mcitedefaultmidpunct}
{\mcitedefaultendpunct}{\mcitedefaultseppunct}\relax
\EndOfBibitem
\bibitem[Best \latin{et~al.}(2012)Best, Zhu, Shim, Lopes, Mittal, Feig, and
  {MacKerell, Jr.}]{BEST2012}
Best,~R.~B.; Zhu,~X.; Shim,~J.; Lopes,~P. E.~M.; Mittal,~J.; Feig,~M.;
  {MacKerell, Jr.},~A.~D. \emph{J.~Chem.\ Theory\ Comput.} \textbf{2012},
  \emph{8}, 3257--3273\relax
\mciteBstWouldAddEndPuncttrue
\mciteSetBstMidEndSepPunct{\mcitedefaultmidpunct}
{\mcitedefaultendpunct}{\mcitedefaultseppunct}\relax
\EndOfBibitem
\bibitem[Lindorff-Larsen \latin{et~al.}(2010)Lindorff-Larsen, Piana, Palmo,
  Maragakis, Klepeis, Dror, and Shaw]{LIND2010B}
Lindorff-Larsen,~K.; Piana,~S.; Palmo,~K.; Maragakis,~P.; Klepeis,~J.~L.;
  Dror,~R.~O.; Shaw,~D.~E. \emph{Proteins:\ Struct.,\ Func.,\ Bioinf.}
  \textbf{2010}, \emph{78}, 1950--8\relax
\mciteBstWouldAddEndPuncttrue
\mciteSetBstMidEndSepPunct{\mcitedefaultmidpunct}
{\mcitedefaultendpunct}{\mcitedefaultseppunct}\relax
\EndOfBibitem
\bibitem[Klauda \latin{et~al.}(2005)Klauda, Brooks, {MacKerell, Jr.}, Venable,
  and Pastor]{KLAU2005}
Klauda,~J.~B.; Brooks,~B.~R.; {MacKerell, Jr.},~A.~D.; Venable,~R.~M.;
  Pastor,~R.~W. \emph{J.~Phys.\ Chem.~B} \textbf{2005}, \emph{109},
  5300--11\relax
\mciteBstWouldAddEndPuncttrue
\mciteSetBstMidEndSepPunct{\mcitedefaultmidpunct}
{\mcitedefaultendpunct}{\mcitedefaultseppunct}\relax
\EndOfBibitem
\bibitem[Klauda \latin{et~al.}(2010)Klauda, Venable, Freites, O'Connor, Tobias,
  Mondragon-Ramirez, Vorobyov, {MacKerell, Jr.}, and Pastor]{KLAU2010}
Klauda,~J.~B.; Venable,~R.~M.; Freites,~J.~A.; O'Connor,~J.~W.; Tobias,~D.~J.;
  Mondragon-Ramirez,~C.; Vorobyov,~I.; {MacKerell, Jr.},~A.~D.; Pastor,~R.~W.
  \emph{J.~Phys.\ Chem.~B} \textbf{2010}, \emph{114}, 7830--7843\relax
\mciteBstWouldAddEndPuncttrue
\mciteSetBstMidEndSepPunct{\mcitedefaultmidpunct}
{\mcitedefaultendpunct}{\mcitedefaultseppunct}\relax
\EndOfBibitem
\bibitem[{MacKerell, Jr.} and Banavali(2000){MacKerell, Jr.}, and
  Banavali]{CHARMMDNA2}
{MacKerell, Jr.},~A.~D.; Banavali,~N.~K. \emph{J.~Comput.\ Chem.}
  \textbf{2000}, \emph{21}, 105--120\relax
\mciteBstWouldAddEndPuncttrue
\mciteSetBstMidEndSepPunct{\mcitedefaultmidpunct}
{\mcitedefaultendpunct}{\mcitedefaultseppunct}\relax
\EndOfBibitem
\bibitem[Foloppe and {MacKerell, Jr.}(2000)Foloppe, and {MacKerell,
  Jr.}]{FOLO2000}
Foloppe,~N.; {MacKerell, Jr.},~A.~D. \emph{J.~Comput.\ Chem.} \textbf{2000},
  \emph{21}, 86--104\relax
\mciteBstWouldAddEndPuncttrue
\mciteSetBstMidEndSepPunct{\mcitedefaultmidpunct}
{\mcitedefaultendpunct}{\mcitedefaultseppunct}\relax
\EndOfBibitem
\bibitem[Perez \latin{et~al.}(2007)Perez, Marchan, Svozil, Sponer, Cheatham,
  Laughton, and Orozco]{PERE2007}
Perez,~A.; Marchan,~I.; Svozil,~D.; Sponer,~J.; Cheatham,~T.~E.;
  Laughton,~C.~A.; Orozco,~M. \emph{Biophys.~J.} \textbf{2007}, \emph{92},
  3817--3829\relax
\mciteBstWouldAddEndPuncttrue
\mciteSetBstMidEndSepPunct{\mcitedefaultmidpunct}
{\mcitedefaultendpunct}{\mcitedefaultseppunct}\relax
\EndOfBibitem
\bibitem[Hart \latin{et~al.}(2012)Hart, Foloppe, Baker, Denning, Nilsson, and
  {MacKerell, Jr.}]{HART2012}
Hart,~K.; Foloppe,~N.; Baker,~C.~M.; Denning,~E.~J.; Nilsson,~L.; {MacKerell,
  Jr.},~A.~D. \emph{J.~Chem.\ Theory\ Comput.} \textbf{2012}, \emph{8},
  348--362\relax
\mciteBstWouldAddEndPuncttrue
\mciteSetBstMidEndSepPunct{\mcitedefaultmidpunct}
{\mcitedefaultendpunct}{\mcitedefaultseppunct}\relax
\EndOfBibitem
\bibitem[Denning \latin{et~al.}(2011)Denning, Priyakumar, Nilsson, and
  {MacKerell, Jr.}]{DENN2011}
Denning,~E.~J.; Priyakumar,~U.~D.; Nilsson,~L.; {MacKerell, Jr.},~A.~D.
  \emph{J.~Comput.\ Chem.} \textbf{2011}, \emph{32}, 1929--1943\relax
\mciteBstWouldAddEndPuncttrue
\mciteSetBstMidEndSepPunct{\mcitedefaultmidpunct}
{\mcitedefaultendpunct}{\mcitedefaultseppunct}\relax
\EndOfBibitem
\bibitem[Zgarbova \latin{et~al.}(2011)Zgarbova, Otyepka, Sponer, Mladek, Banas,
  Thomas E.~Cheatham, and Jurecka]{ZGAR2011B}
Zgarbova,~M.; Otyepka,~M.; Sponer,~J.; Mladek,~A.; Banas,~P.; Thomas
  E.~Cheatham,~I.; Jurecka,~P. \emph{J.~Chem.\ Theory\ Comput.} \textbf{2011},
  \emph{7}, 2886--2902, PMID: 21921995\relax
\mciteBstWouldAddEndPuncttrue
\mciteSetBstMidEndSepPunct{\mcitedefaultmidpunct}
{\mcitedefaultendpunct}{\mcitedefaultseppunct}\relax
\EndOfBibitem
\bibitem[Fadrna \latin{et~al.}(2009)Fadrna, Spackova, Sarzynska, Koca, Orozco,
  Thomas E.~Cheatham, Kulinski, and Sponer]{FADR2009}
Fadrna,~E.; Spackova,~N.; Sarzynska,~J.; Koca,~J.; Orozco,~M.; Thomas
  E.~Cheatham,~I.; Kulinski,~T.; Sponer,~J. \emph{J.~Chem.\ Theory\ Comput.}
  \textbf{2009}, \emph{5}, 2514--2530\relax
\mciteBstWouldAddEndPuncttrue
\mciteSetBstMidEndSepPunct{\mcitedefaultmidpunct}
{\mcitedefaultendpunct}{\mcitedefaultseppunct}\relax
\EndOfBibitem
\bibitem[{MacKerell, Jr.} \latin{et~al.}(1998){MacKerell, Jr.}, Bashford,
  Bellott, {Dunbrack, Jr.}, Evanseck, Field, Fischer, Gao, Guo, Ha,
  Joseph-{McCarthy}, Kuchnir, Kuczera, Lau, Mattos, Michnick, Ngo, Nguyen,
  Prodhom, {Reiher, III}, Roux, Schlenkrich, Smith, Stote, Straub, Watanabe,
  Wi\'{o}rkiewicz-Kuczera, Yin, and Karplus]{MACK1998}
{MacKerell, Jr.},~A.~D.; Bashford,~D.; Bellott,~M.; {Dunbrack, Jr.},~R.~L.;
  Evanseck,~J.~D.; Field,~M.~J.; Fischer,~S.; Gao,~J.; Guo,~H.; Ha,~S.;
  Joseph-{McCarthy},~D.; Kuchnir,~L.; Kuczera,~K.; Lau,~F. T.~K.; Mattos,~C.;
  Michnick,~S.; Ngo,~T.; Nguyen,~D.~T.; Prodhom,~B.; {Reiher, III},~W.~E.;
  Roux,~B.; Schlenkrich,~M.; Smith,~J.~C.; Stote,~R.; Straub,~J.; Watanabe,~M.;
  Wi\'{o}rkiewicz-Kuczera,~J.; Yin,~D.; Karplus,~M. \emph{J.~Phys.\ Chem.~B}
  \textbf{1998}, \emph{102}, 3586--3616\relax
\mciteBstWouldAddEndPuncttrue
\mciteSetBstMidEndSepPunct{\mcitedefaultmidpunct}
{\mcitedefaultendpunct}{\mcitedefaultseppunct}\relax
\EndOfBibitem
\bibitem[Petrov and Zagrovic(2014)Petrov, and Zagrovic]{PETR2014}
Petrov,~D.; Zagrovic,~B. \emph{PLoS Comput.\ Biol.} \textbf{2014}, \emph{10},
  e1003638\relax
\mciteBstWouldAddEndPuncttrue
\mciteSetBstMidEndSepPunct{\mcitedefaultmidpunct}
{\mcitedefaultendpunct}{\mcitedefaultseppunct}\relax
\EndOfBibitem
\bibitem[Johnson \latin{et~al.}(2009)Johnson, Malardier-Jugroot, Murarka, and
  Head-Gordon]{JOHN2009}
Johnson,~M.~E.; Malardier-Jugroot,~C.; Murarka,~R.~K.; Head-Gordon,~T.
  \emph{J.~Phys.\ Chem.~B} \textbf{2009}, \emph{113}, 4082--4092\relax
\mciteBstWouldAddEndPuncttrue
\mciteSetBstMidEndSepPunct{\mcitedefaultmidpunct}
{\mcitedefaultendpunct}{\mcitedefaultseppunct}\relax
\EndOfBibitem
\bibitem[Nerenberg \latin{et~al.}(2012)Nerenberg, Jo, So, Tripathy, and
  Head-Gordon]{NERE2012}
Nerenberg,~P.~S.; Jo,~B.; So,~C.; Tripathy,~A.; Head-Gordon,~T. \emph{J.~Phys.\
  Chem.~B} \textbf{2012}, \emph{116}, 4524--4534\relax
\mciteBstWouldAddEndPuncttrue
\mciteSetBstMidEndSepPunct{\mcitedefaultmidpunct}
{\mcitedefaultendpunct}{\mcitedefaultseppunct}\relax
\EndOfBibitem
\bibitem[Best \latin{et~al.}(2014)Best, Zheng, and Mittal]{BEST2014}
Best,~R.~B.; Zheng,~W.; Mittal,~J. \emph{J.~Chem.\ Theory\ Comput.}
  \textbf{2014}, \emph{10}, 5113--5124\relax
\mciteBstWouldAddEndPuncttrue
\mciteSetBstMidEndSepPunct{\mcitedefaultmidpunct}
{\mcitedefaultendpunct}{\mcitedefaultseppunct}\relax
\EndOfBibitem
\bibitem[Piana \latin{et~al.}(2015)Piana, Donchev, Robustelli, and
  Shaw]{PIAN2015}
Piana,~S.; Donchev,~A.~G.; Robustelli,~P.; Shaw,~D.~E. \emph{J.~Phys.\ Chem.~B}
  \textbf{2015}, \emph{119}, 5113--5123\relax
\mciteBstWouldAddEndPuncttrue
\mciteSetBstMidEndSepPunct{\mcitedefaultmidpunct}
{\mcitedefaultendpunct}{\mcitedefaultseppunct}\relax
\EndOfBibitem
\bibitem[Yoo and Aksimentiev(2012)Yoo, and Aksimentiev]{YOO2012}
Yoo,~J.; Aksimentiev,~A. \emph{J.~Phys.\ Chem.\ Lett.} \textbf{2012}, \emph{3},
  45--50\relax
\mciteBstWouldAddEndPuncttrue
\mciteSetBstMidEndSepPunct{\mcitedefaultmidpunct}
{\mcitedefaultendpunct}{\mcitedefaultseppunct}\relax
\EndOfBibitem
\bibitem[Rau \latin{et~al.}(1984)Rau, Lee, and Parsegian]{RAU1984}
Rau,~D.~C.; Lee,~B.; Parsegian,~V.~A. \emph{Proc.\ Natl.\ Acad.\ Sci.\ U.S.A.}
  \textbf{1984}, \emph{81}, 2621--2625\relax
\mciteBstWouldAddEndPuncttrue
\mciteSetBstMidEndSepPunct{\mcitedefaultmidpunct}
{\mcitedefaultendpunct}{\mcitedefaultseppunct}\relax
\EndOfBibitem
\bibitem[Luo and Roux(2009)Luo, and Roux]{LUO2009b}
Luo,~Y.; Roux,~B. \emph{J.~Phys.\ Chem.\ Lett.} \textbf{2009}, \emph{1},
  183--9\relax
\mciteBstWouldAddEndPuncttrue
\mciteSetBstMidEndSepPunct{\mcitedefaultmidpunct}
{\mcitedefaultendpunct}{\mcitedefaultseppunct}\relax
\EndOfBibitem
\bibitem[Yoo and Aksimentiev(2012)Yoo, and Aksimentiev]{YOO2012B}
Yoo,~J.; Aksimentiev,~A. \emph{J.~Phys.\ Chem.~B} \textbf{2012}, \emph{116},
  12946--12954\relax
\mciteBstWouldAddEndPuncttrue
\mciteSetBstMidEndSepPunct{\mcitedefaultmidpunct}
{\mcitedefaultendpunct}{\mcitedefaultseppunct}\relax
\EndOfBibitem
\bibitem[Maffeo \latin{et~al.}(2014)Maffeo, Ngo, Ha, and Aksimentiev]{MAFF2014}
Maffeo,~C.; Ngo,~T. T.~M.; Ha,~T.; Aksimentiev,~A. \emph{J.~Chem.\ Theory\
  Comput.} \textbf{2014}, \emph{10}, 2891--2896\relax
\mciteBstWouldAddEndPuncttrue
\mciteSetBstMidEndSepPunct{\mcitedefaultmidpunct}
{\mcitedefaultendpunct}{\mcitedefaultseppunct}\relax
\EndOfBibitem
\bibitem[Yoo and Aksimentiev()Yoo, and Aksimentiev]{YOO2015A}
Yoo,~J.; Aksimentiev,~A. Submitted\relax
\mciteBstWouldAddEndPuncttrue
\mciteSetBstMidEndSepPunct{\mcitedefaultmidpunct}
{\mcitedefaultendpunct}{\mcitedefaultseppunct}\relax
\EndOfBibitem
\bibitem[Jorgensen \latin{et~al.}(1983)Jorgensen, Chandrasekhar, Madura, Impey,
  and Klein]{JORG1983}
Jorgensen,~W.~L.; Chandrasekhar,~J.; Madura,~J.~D.; Impey,~R.~W.; Klein,~M.~L.
  \emph{J.~Chem.\ Phys.} \textbf{1983}, \emph{79}, 926--935\relax
\mciteBstWouldAddEndPuncttrue
\mciteSetBstMidEndSepPunct{\mcitedefaultmidpunct}
{\mcitedefaultendpunct}{\mcitedefaultseppunct}\relax
\EndOfBibitem
\bibitem[Price and {Brooks III}(2004)Price, and {Brooks III}]{PRIC2004}
Price,~D.~J.; {Brooks III},~C.~L. \emph{J.~Chem.\ Phys.} \textbf{2004},
  \emph{121}, 10096\relax
\mciteBstWouldAddEndPuncttrue
\mciteSetBstMidEndSepPunct{\mcitedefaultmidpunct}
{\mcitedefaultendpunct}{\mcitedefaultseppunct}\relax
\EndOfBibitem
\bibitem[Beglov and Roux(1994)Beglov, and Roux]{BEGL1994}
Beglov,~D.; Roux,~B. \emph{J.~Chem.\ Phys.} \textbf{1994}, \emph{100},
  9050--9063\relax
\mciteBstWouldAddEndPuncttrue
\mciteSetBstMidEndSepPunct{\mcitedefaultmidpunct}
{\mcitedefaultendpunct}{\mcitedefaultseppunct}\relax
\EndOfBibitem
\bibitem[Cannon \latin{et~al.}(1994)Cannon, Pettitt, and McCammon]{CANN1994}
Cannon,~W.~R.; Pettitt,~B.~M.; McCammon,~J.~A. \emph{J.~Phys.\ Chem.}
  \textbf{1994}, \emph{98}, 6225--6230\relax
\mciteBstWouldAddEndPuncttrue
\mciteSetBstMidEndSepPunct{\mcitedefaultmidpunct}
{\mcitedefaultendpunct}{\mcitedefaultseppunct}\relax
\EndOfBibitem
\bibitem[Ullmann \latin{et~al.}(2012)Ullmann, Andrade, and Ullmann]{ULLM2012}
Ullmann,~R.~T.; Andrade,~S. L.~A.; Ullmann,~G.~M. \emph{J.~Phys.\ Chem.~B}
  \textbf{2012}, \emph{116}, 9690--9703\relax
\mciteBstWouldAddEndPuncttrue
\mciteSetBstMidEndSepPunct{\mcitedefaultmidpunct}
{\mcitedefaultendpunct}{\mcitedefaultseppunct}\relax
\EndOfBibitem
\bibitem[Lindstr{\"o}m and Andersson-Svahn(2010)Lindstr{\"o}m, and
  Andersson-Svahn]{LIND2010}
Lindstr{\"o}m,~S.; Andersson-Svahn,~H. \emph{Lab Chip} \textbf{2010},
  \emph{10}, 3363--3372\relax
\mciteBstWouldAddEndPuncttrue
\mciteSetBstMidEndSepPunct{\mcitedefaultmidpunct}
{\mcitedefaultendpunct}{\mcitedefaultseppunct}\relax
\EndOfBibitem
\bibitem[Nerenberg and Head-Gordon(2011)Nerenberg, and Head-Gordon]{NERE2011}
Nerenberg,~P.~S.; Head-Gordon,~T. \emph{J.~Chem.\ Theory\ Comput.}
  \textbf{2011}, \emph{7}, 1220--1230\relax
\mciteBstWouldAddEndPuncttrue
\mciteSetBstMidEndSepPunct{\mcitedefaultmidpunct}
{\mcitedefaultendpunct}{\mcitedefaultseppunct}\relax
\EndOfBibitem
\bibitem[Beauchamp \latin{et~al.}(2012)Beauchamp, Lin, Das, and
  Pande]{BEAU2012}
Beauchamp,~K.~A.; Lin,~Y.-S. .~S.; Das,~R.; Pande,~V.~S. \emph{J.~Chem.\
  Theory\ Comput.} \textbf{2012}, \emph{8}, 1409--1414\relax
\mciteBstWouldAddEndPuncttrue
\mciteSetBstMidEndSepPunct{\mcitedefaultmidpunct}
{\mcitedefaultendpunct}{\mcitedefaultseppunct}\relax
\EndOfBibitem
\bibitem[Joung and Cheatham(2008)Joung, and Cheatham]{JOUN2008}
Joung,~I.~S.; Cheatham,~T.~E. \emph{J.~Phys.\ Chem.~B} \textbf{2008},
  \emph{112}, 9020--9041\relax
\mciteBstWouldAddEndPuncttrue
\mciteSetBstMidEndSepPunct{\mcitedefaultmidpunct}
{\mcitedefaultendpunct}{\mcitedefaultseppunct}\relax
\EndOfBibitem
\bibitem[Phillips \latin{et~al.}(2005)Phillips, Braun, Wang, Gumbart,
  Tajkhorshid, Villa, Chipot, Skeel, Kale, and Schulten]{PHIL2005}
Phillips,~J.~C.; Braun,~R.; Wang,~W.; Gumbart,~J.; Tajkhorshid,~E.; Villa,~E.;
  Chipot,~C.; Skeel,~R.~D.; Kale,~L.; Schulten,~K. \emph{J.~Comput.\ Chem.}
  \textbf{2005}, \emph{26}, 1781--1802\relax
\mciteBstWouldAddEndPuncttrue
\mciteSetBstMidEndSepPunct{\mcitedefaultmidpunct}
{\mcitedefaultendpunct}{\mcitedefaultseppunct}\relax
\EndOfBibitem
\bibitem[Luo \latin{et~al.}(2010)Luo, Egwolf, Walters, and Roux]{LUO2009}
Luo,~Y.; Egwolf,~B.; Walters,~D.~E.; Roux,~B. \emph{J.~Phys.\ Chem.~B}
  \textbf{2010}, \emph{114}, 952--958\relax
\mciteBstWouldAddEndPuncttrue
\mciteSetBstMidEndSepPunct{\mcitedefaultmidpunct}
{\mcitedefaultendpunct}{\mcitedefaultseppunct}\relax
\EndOfBibitem
\bibitem[Heng \latin{et~al.}(2006)Heng, Aksimentiev, Ho, Marks, Grinkova,
  Sligar, Schulten, and Timp]{HENG2006}
Heng,~J.~B.; Aksimentiev,~A.; Ho,~C.; Marks,~P.; Grinkova,~Y.~V.; Sligar,~S.;
  Schulten,~K.; Timp,~G. \emph{Biophys.~J.} \textbf{2006}, \emph{90},
  1098--1106\relax
\mciteBstWouldAddEndPuncttrue
\mciteSetBstMidEndSepPunct{\mcitedefaultmidpunct}
{\mcitedefaultendpunct}{\mcitedefaultseppunct}\relax
\EndOfBibitem
\bibitem[Koopman and Lowe(2006)Koopman, and Lowe]{KOOP2006}
Koopman,~E.~A.; Lowe,~C.~P. \emph{J.~Chem.\ Phys.} \textbf{2006}, \emph{124},
  --\relax
\mciteBstWouldAddEndPuncttrue
\mciteSetBstMidEndSepPunct{\mcitedefaultmidpunct}
{\mcitedefaultendpunct}{\mcitedefaultseppunct}\relax
\EndOfBibitem
\bibitem[Hess \latin{et~al.}(2008)Hess, Kutzner, {van der Spoel}, and
  Lindahl]{HESS2008}
Hess,~B.; Kutzner,~C.; {van der Spoel},~D.; Lindahl,~E. \emph{J.~Chem.\ Theory\
  Comput.} \textbf{2008}, \emph{4}, 435--447\relax
\mciteBstWouldAddEndPuncttrue
\mciteSetBstMidEndSepPunct{\mcitedefaultmidpunct}
{\mcitedefaultendpunct}{\mcitedefaultseppunct}\relax
\EndOfBibitem
\bibitem[Nose and Klein(1983)Nose, and Klein]{NOSE1983}
Nose,~S.; Klein,~M.~L. \emph{Mol.\ Phys.} \textbf{1983}, \emph{50},
  1055--76\relax
\mciteBstWouldAddEndPuncttrue
\mciteSetBstMidEndSepPunct{\mcitedefaultmidpunct}
{\mcitedefaultendpunct}{\mcitedefaultseppunct}\relax
\EndOfBibitem
\bibitem[Hoover(1985)]{HOOV1985}
Hoover,~W.~G. \emph{Phys.\ Rev.~A} \textbf{1985}, \emph{31}, 1695--1697\relax
\mciteBstWouldAddEndPuncttrue
\mciteSetBstMidEndSepPunct{\mcitedefaultmidpunct}
{\mcitedefaultendpunct}{\mcitedefaultseppunct}\relax
\EndOfBibitem
\bibitem[Parrinello and Rahman(1981)Parrinello, and Rahman]{PARR1981}
Parrinello,~M.; Rahman,~A. \emph{J.~ Appl.\ Phys.} \textbf{1981}, \emph{52},
  7182--90\relax
\mciteBstWouldAddEndPuncttrue
\mciteSetBstMidEndSepPunct{\mcitedefaultmidpunct}
{\mcitedefaultendpunct}{\mcitedefaultseppunct}\relax
\EndOfBibitem
\bibitem[Darden \latin{et~al.}(1993)Darden, York, and Pedersen]{DARD1993}
Darden,~T.~A.; York,~D.; Pedersen,~L. \emph{J.~Chem.\ Phys.} \textbf{1993},
  \emph{98}, 10089--92\relax
\mciteBstWouldAddEndPuncttrue
\mciteSetBstMidEndSepPunct{\mcitedefaultmidpunct}
{\mcitedefaultendpunct}{\mcitedefaultseppunct}\relax
\EndOfBibitem
\bibitem[Miyamoto and Kollman(1992)Miyamoto, and Kollman]{MIYA1992}
Miyamoto,~S.; Kollman,~P.~A. \emph{J.~Comput.\ Chem.} \textbf{1992}, \emph{13},
  952--62\relax
\mciteBstWouldAddEndPuncttrue
\mciteSetBstMidEndSepPunct{\mcitedefaultmidpunct}
{\mcitedefaultendpunct}{\mcitedefaultseppunct}\relax
\EndOfBibitem
\bibitem[Hess \latin{et~al.}(1997)Hess, Bekker, Berendsen, and
  Fraaije]{HESS1997}
Hess,~B.; Bekker,~H.; Berendsen,~H. J.~C.; Fraaije,~J. G. E.~M.
  \emph{J.~Comput.\ Chem.} \textbf{1997}, \emph{18}, 1463--72\relax
\mciteBstWouldAddEndPuncttrue
\mciteSetBstMidEndSepPunct{\mcitedefaultmidpunct}
{\mcitedefaultendpunct}{\mcitedefaultseppunct}\relax
\EndOfBibitem
\bibitem[Todd \latin{et~al.}(2008)Todd, Parsegian, Shirahata, Thomas, and
  Rau]{TODD2008}
Todd,~B.~A.; Parsegian,~V.~A.; Shirahata,~A.; Thomas,~T.~J.; Rau,~D.~C.
  \emph{Biophys.~J.} \textbf{2008}, \emph{94}, 4775--82\relax
\mciteBstWouldAddEndPuncttrue
\mciteSetBstMidEndSepPunct{\mcitedefaultmidpunct}
{\mcitedefaultendpunct}{\mcitedefaultseppunct}\relax
\EndOfBibitem
\bibitem[Kumar \latin{et~al.}(1992)Kumar, Rosenberg, Bouzida, Swendsen, and
  Kollman]{KUMA1992}
Kumar,~S.; Rosenberg,~J.~M.; Bouzida,~D.; Swendsen,~R.~H.; Kollman,~P.~A.
  \emph{J.~Comput.\ Chem.} \textbf{1992}, \emph{13}, 1011--1021\relax
\mciteBstWouldAddEndPuncttrue
\mciteSetBstMidEndSepPunct{\mcitedefaultmidpunct}
{\mcitedefaultendpunct}{\mcitedefaultseppunct}\relax
\EndOfBibitem
\bibitem[Brooks \latin{et~al.}(2009)Brooks, Brooks, {MacKerell, Jr.}, Nilsson,
  Petrella, Roux, Won, Archontis, Bartels, Boresch, Caflisch, Caves, Cui,
  Dinner, Feig, Fischer, Gao, Hodoscek, Im, Kuczera, Lazaridis, Ma,
  Ovchinnikov, Paci, Pastor, Post, Pu, Schaefer, Tidor, Venable, Woodcock, Wu,
  Yang, York, and Karplus]{BROO2009b}
Brooks,~B.~R.; Brooks,~C.~L.; {MacKerell, Jr.},~A.~D.; Nilsson,~L.;
  Petrella,~R.~J.; Roux,~B.; Won,~Y.; Archontis,~G.; Bartels,~C.; Boresch,~S.;
  Caflisch,~A.; Caves,~L.; Cui,~Q.; Dinner,~A.~R.; Feig,~M.; Fischer,~S.;
  Gao,~J.; Hodoscek,~M.; Im,~W.; Kuczera,~K.; Lazaridis,~T.; Ma,~J.;
  Ovchinnikov,~V.; Paci,~E.; Pastor,~R.~W.; Post,~C.~B.; Pu,~J.~Z.;
  Schaefer,~M.; Tidor,~B.; Venable,~R.~M.; Woodcock,~H.~L.; Wu,~X.; Yang,~W.;
  York,~D.~M.; Karplus,~M. \emph{J.~Comput.\ Chem.} \textbf{2009}, \emph{30},
  1545--614\relax
\mciteBstWouldAddEndPuncttrue
\mciteSetBstMidEndSepPunct{\mcitedefaultmidpunct}
{\mcitedefaultendpunct}{\mcitedefaultseppunct}\relax
\EndOfBibitem
\bibitem[Hynninen and Crowley(2014)Hynninen, and Crowley]{HYNN2014}
Hynninen,~A.-P.; Crowley,~M.~F. \emph{J.~Comput.\ Chem.} \textbf{2014},
  \emph{35}, 406--413\relax
\mciteBstWouldAddEndPuncttrue
\mciteSetBstMidEndSepPunct{\mcitedefaultmidpunct}
{\mcitedefaultendpunct}{\mcitedefaultseppunct}\relax
\EndOfBibitem
\bibitem[Klauda()]{KLAUDA}
Klauda,~J.~B.
  \emph{http://terpconnect.umd.edu/$\sim$jbklauda/research/download.html}
  \relax
\mciteBstWouldAddEndPunctfalse
\mciteSetBstMidEndSepPunct{\mcitedefaultmidpunct}
{}{\mcitedefaultseppunct}\relax
\EndOfBibitem
\bibitem[Tsurko \latin{et~al.}(2007)Tsurko, Neueder, and Kunz]{TSUR2007}
Tsurko,~E.; Neueder,~R.; Kunz,~W. \emph{J.~Solut.\ Chem.} \textbf{2007},
  \emph{36}, 651--672\relax
\mciteBstWouldAddEndPuncttrue
\mciteSetBstMidEndSepPunct{\mcitedefaultmidpunct}
{\mcitedefaultendpunct}{\mcitedefaultseppunct}\relax
\EndOfBibitem
\bibitem[Masunov and Lazaridis(2003)Masunov, and Lazaridis]{MASU2003}
Masunov,~A.; Lazaridis,~T. \emph{J.~Am.\ Chem.\ Soc.} \textbf{2003},
  \emph{125}, 1722--30\relax
\mciteBstWouldAddEndPuncttrue
\mciteSetBstMidEndSepPunct{\mcitedefaultmidpunct}
{\mcitedefaultendpunct}{\mcitedefaultseppunct}\relax
\EndOfBibitem
\bibitem[Lide(2005)]{crc_handbook}
Lide,~D.~R., Ed. \emph{{CRC} Handbook Chemistry and Physics}, 85th ed.; CRC
  Press, 2005\relax
\mciteBstWouldAddEndPuncttrue
\mciteSetBstMidEndSepPunct{\mcitedefaultmidpunct}
{\mcitedefaultendpunct}{\mcitedefaultseppunct}\relax
\EndOfBibitem
\bibitem[Robinson and Stokes(1959)Robinson, and Stokes]{ROBI1959}
Robinson,~R.~A.; Stokes,~R.~H. \emph{Electrolyte solutions}; Butterworths
  scientific publications, 1959\relax
\mciteBstWouldAddEndPuncttrue
\mciteSetBstMidEndSepPunct{\mcitedefaultmidpunct}
{\mcitedefaultendpunct}{\mcitedefaultseppunct}\relax
\EndOfBibitem
\bibitem[Cleland and Hengge(2006)Cleland, and Hengge]{CLEL2006}
Cleland,~W.~W.; Hengge,~A.~C. \emph{Chem.\ Rev.} \textbf{2006}, \emph{106},
  3252--78\relax
\mciteBstWouldAddEndPuncttrue
\mciteSetBstMidEndSepPunct{\mcitedefaultmidpunct}
{\mcitedefaultendpunct}{\mcitedefaultseppunct}\relax
\EndOfBibitem
\bibitem[L\"{u}bben \latin{et~al.}(2007)L\"{u}bben, G\"{u}ldenhaupt, Zoltner,
  Deigweiher, Haebel, Urbanke, and Scheidig]{LUBB2007}
L\"{u}bben,~M.; G\"{u}ldenhaupt,~J.; Zoltner,~M.; Deigweiher,~K.; Haebel,~P.;
  Urbanke,~C.; Scheidig,~A.~J. \emph{J.~Mol.\ Biol.} \textbf{2007}, \emph{369},
  368--85\relax
\mciteBstWouldAddEndPuncttrue
\mciteSetBstMidEndSepPunct{\mcitedefaultmidpunct}
{\mcitedefaultendpunct}{\mcitedefaultseppunct}\relax
\EndOfBibitem
\bibitem[Copley and Barton(1994)Copley, and Barton]{COPL1994}
Copley,~R.~R.; Barton,~G.~J. \emph{J.~Mol.\ Biol.} \textbf{1994}, \emph{242},
  321 -- 329\relax
\mciteBstWouldAddEndPuncttrue
\mciteSetBstMidEndSepPunct{\mcitedefaultmidpunct}
{\mcitedefaultendpunct}{\mcitedefaultseppunct}\relax
\EndOfBibitem
\bibitem[Omi \latin{et~al.}(2007)Omi, Goto, Miyahara, Manzoku, Ebihara, and
  Hirotsu]{RIE2007}
Omi,~R.; Goto,~M.; Miyahara,~I.; Manzoku,~M.; Ebihara,~A.; Hirotsu,~K.
  \emph{Biochemistry} \textbf{2007}, \emph{46}, 12618--12627, PMID:
  17929834\relax
\mciteBstWouldAddEndPuncttrue
\mciteSetBstMidEndSepPunct{\mcitedefaultmidpunct}
{\mcitedefaultendpunct}{\mcitedefaultseppunct}\relax
\EndOfBibitem
\bibitem[Kumar \latin{et~al.}(2010)Kumar, Singh, Gautam, and
  Karthikeyan]{KUMA2010}
Kumar,~P.; Singh,~M.; Gautam,~R.; Karthikeyan,~S. \emph{Proteins:\ Struct.,\
  Func.,\ Bioinf.} \textbf{2010}, \emph{78}, 3292--303\relax
\mciteBstWouldAddEndPuncttrue
\mciteSetBstMidEndSepPunct{\mcitedefaultmidpunct}
{\mcitedefaultendpunct}{\mcitedefaultseppunct}\relax
\EndOfBibitem
\bibitem[Ghodge \latin{et~al.}(2013)Ghodge, Fedorov, Fedorov, Hillerich,
  Seidel, Almo, and Raushel]{SWAP2013}
Ghodge,~S.~V.; Fedorov,~A.~A.; Fedorov,~E.~V.; Hillerich,~B.; Seidel,~R.;
  Almo,~S.~C.; Raushel,~F.~M. \emph{Biochemistry} \textbf{2013}, \emph{52},
  1101--1112\relax
\mciteBstWouldAddEndPuncttrue
\mciteSetBstMidEndSepPunct{\mcitedefaultmidpunct}
{\mcitedefaultendpunct}{\mcitedefaultseppunct}\relax
\EndOfBibitem
\bibitem[Dai \latin{et~al.}(2008)Dai, Mu, Nordenski{\"o}ld, and {van der
  Maarel}]{DAI2008}
Dai,~L.; Mu,~Y.; Nordenski{\"o}ld,~L.; {van der Maarel},~J. R.~C. \emph{Phys.\
  Rev.\ Lett.} \textbf{2008}, \emph{100}, 118301\relax
\mciteBstWouldAddEndPuncttrue
\mciteSetBstMidEndSepPunct{\mcitedefaultmidpunct}
{\mcitedefaultendpunct}{\mcitedefaultseppunct}\relax
\EndOfBibitem
\bibitem[DeRouchey \latin{et~al.}(2013)DeRouchey, Hoover, and Rau]{DERO2013}
DeRouchey,~J.; Hoover,~B.; Rau,~D.~C. \emph{Biochemistry} \textbf{2013},
  \emph{52}, 3000--3009\relax
\mciteBstWouldAddEndPuncttrue
\mciteSetBstMidEndSepPunct{\mcitedefaultmidpunct}
{\mcitedefaultendpunct}{\mcitedefaultseppunct}\relax
\EndOfBibitem
\bibitem[Maffeo \latin{et~al.}(2012)Maffeo, Luan, and Aksimentiev]{MAFF2012A}
Maffeo,~C.; Luan,~B.; Aksimentiev,~A. \emph{Nucl.\ Acids\ Res.} \textbf{2012},
  \emph{40}, 3812--3821\relax
\mciteBstWouldAddEndPuncttrue
\mciteSetBstMidEndSepPunct{\mcitedefaultmidpunct}
{\mcitedefaultendpunct}{\mcitedefaultseppunct}\relax
\EndOfBibitem
\bibitem[Luan and Aksimentiev(2008)Luan, and Aksimentiev]{LUAN2008C}
Luan,~B.; Aksimentiev,~A. \emph{J.~Am.\ Chem.\ Soc.} \textbf{2008}, \emph{130},
  15754--15755\relax
\mciteBstWouldAddEndPuncttrue
\mciteSetBstMidEndSepPunct{\mcitedefaultmidpunct}
{\mcitedefaultendpunct}{\mcitedefaultseppunct}\relax
\EndOfBibitem
\bibitem[Maffeo \latin{et~al.}(2010)Maffeo, Sch{\"o}pflin, Brutzer, Stehr,
  Aksimentiev, Wedemann, and Seidel]{MAFF2010B}
Maffeo,~C.; Sch{\"o}pflin,~R.; Brutzer,~H.; Stehr,~R.; Aksimentiev,~A.;
  Wedemann,~G.; Seidel,~R. \emph{Phys.\ Rev.\ Lett.} \textbf{2010}, \emph{105},
  158101\relax
\mciteBstWouldAddEndPuncttrue
\mciteSetBstMidEndSepPunct{\mcitedefaultmidpunct}
{\mcitedefaultendpunct}{\mcitedefaultseppunct}\relax
\EndOfBibitem
\bibitem[Petrache \latin{et~al.}(2000)Petrache, Dodd, and Brown]{PETR2000}
Petrache,~H.~I.; Dodd,~S.~W.; Brown,~M.~F. \emph{Biophys.~J.} \textbf{2000},
  \emph{79}, 3172--92\relax
\mciteBstWouldAddEndPuncttrue
\mciteSetBstMidEndSepPunct{\mcitedefaultmidpunct}
{\mcitedefaultendpunct}{\mcitedefaultseppunct}\relax
\EndOfBibitem
\bibitem[Petrache \latin{et~al.}(2004)Petrache, Tristram-Nagle, Gawrisch,
  Harries, Parsegian, and Nagle]{PETR2004}
Petrache,~H.~I.; Tristram-Nagle,~S.; Gawrisch,~K.; Harries,~D.;
  Parsegian,~V.~A.; Nagle,~J.~F. Structure and Fluctuations of Charged
  Phosphatidylserine Bilayers in the Absence of Salt. 2004\relax
\mciteBstWouldAddEndPuncttrue
\mciteSetBstMidEndSepPunct{\mcitedefaultmidpunct}
{\mcitedefaultendpunct}{\mcitedefaultseppunct}\relax
\EndOfBibitem
\bibitem[Klauda \latin{et~al.}(2008)Klauda, Venable, {MacKerell, Jr.}, and
  Pastor]{KLAU2008}
Klauda,~J.~B.; Venable,~R.~M.; {MacKerell, Jr.},~A.~D.; Pastor,~R.~W. In
  \emph{Current Topics in Membranes}; Feller,~S.~E., Ed.; Elsevier, 2008;
  Vol.~60; pp 1--48\relax
\mciteBstWouldAddEndPuncttrue
\mciteSetBstMidEndSepPunct{\mcitedefaultmidpunct}
{\mcitedefaultendpunct}{\mcitedefaultseppunct}\relax
\EndOfBibitem
\bibitem[Brown and London(1998)Brown, and London]{BROW1998}
Brown,~D.~A.; London,~E. \emph{Annual Review of Cell and Developmental Biology}
  \textbf{1998}, \emph{14}, 111--136\relax
\mciteBstWouldAddEndPuncttrue
\mciteSetBstMidEndSepPunct{\mcitedefaultmidpunct}
{\mcitedefaultendpunct}{\mcitedefaultseppunct}\relax
\EndOfBibitem
\end{mcitethebibliography}

\end{document}